\documentclass{aa}

\usepackage{times}
\usepackage{graphicx}
\usepackage{xspace}
\usepackage{epsfig}
\usepackage{natbib}
\usepackage{deluxetable}
\usepackage{mydefs}

\newcommand{\myrule}{\rule[-0.2cm]{0.cm}{0.5cm}}

\newcommand\arcdeg{\mbox{$^\circ$}}

\begin{document}

\title{X-ray view of IC\,348 in the light of an updated cluster census}


\author{B. Stelzer \inst{1} \and T. Preibisch \inst{2} \and F. Alexander \inst{2} \and P. Mucciarelli \inst{2} \and E. Flaccomio \inst{1} \and G. Micela \inst{1} \and S. Sciortino \inst{1}}

\offprints{B. Stelzer}

\institute{INAF - Osservatorio Astronomico di Palermo,
  Piazza del Parlamento 1,
  I-90134 Palermo, Italy \\ \email{B. Stelzer, stelzer@astropa.inaf.it} \and
  Universit\"ats-Sternwarte M\"unchen, Ludwig-Maximilians-Universit\"at,
Scheinerstr.~1, D-81679 M\"unchen, Germany
}


\date{Received $<$17 September 2011$>$ / Accepted $<$8 November 2011$>$}

\abstract
{IC\,348 is a nearby ($\approx 310$\,pc), young ($\sim 2-3$\,Myr) 
open cluster with 
$> 300$\, members identified from optical and infrared observations.  
It comprises Young Stellar Objects in various evolutionary phases from
protostars over disk-bearing to diskless pre-main sequence stars.
This gives us the opportunity to study evolutionary effects in the 
high-energy emissions in a homogeneous environment.
}
{We study the properties of the coronae of the young low-mass stars
in IC\,348 combining X-ray and optical/infrared data. In particular,
we intend to shed light on 
the dependence of X-ray luminosity and spectral hardness
on evolutionary stage and on stellar parameters such as mass, effective
temperature and bolometric luminosity.
}
{The four existing Chandra observations of IC\,348 are merged, thus providing 
a deeper and spatially more complete X-ray view than previous X-ray 
studies of the cluster. We have compiled a comprehensive catalog of
IC\,348 members taking into account recent updates to the cluster census.
Our data collection comprises fundamental stellar parameters, infrared
excess indicating the presence of disks, H$\alpha$ emission as a tracer
of chromospheric emission or accretion and mass accretion rates. 
}
{We have detected $290$ X-ray sources in four merged Chandra exposures,
of which $187$ are associated with known cluster members
corresponding to a detection rate of $\sim 60$\,\% for the cluster members of IC\,348
identified in optical/infrared studies. 
According to the most recent spectral classification of IC\,348 members
only four of the X-ray sources are brown dwarfs (spectral type M6 and later). 
The detection rate is highest for diskless Class\,III stars and increases
with stellar mass. This may be explained with higher X-ray 
luminosities for higher mass and later evolutionary stage that is evident 
in the X-ray luminosity functions. In particular, we find that for the lowest
examined masses ($0.1-0.25\,M_\odot$) there is a difference between
the X-ray luminosity functions of 
accreting and non-accreting stars (classified on the basis of their
H$\alpha$ emission strength) as well as those of 
disk-bearing and diskless stars (classified on the basis of the slope
of the spectral energy distribution). These differences disappear for
higher masses.  
This is related to our finding that the $L_{\rm x}/L_{\rm bol}$ ratio is non-constant
across the mass/luminosity sequence of IC\,348 with a
decrease towards lower luminosity stars. 
Our analysis of an analogous stellar 
sample in the Orion Nebula Cluster suggests that the decline of $L_{\rm x}/L_{\rm bol}$ 
for young stars at the low-mass end of the stellar sequence is likely universal.
} 
{}


\maketitle

\section{Introduction}\label{sect:intro}

The open cluster IC\,348 is arguably the best investigated rich  ($> 100$ members),
and very young ($\sim 2-3$\,Myr old) stellar cluster
within $\sim 300$\,pc from the Sun. IC~348 is associated to 
the Perseus molecular cloud complex at a distance of about $310$\,pc \citep{Herbig98.1} and
located just to the south of the optically bright star $o$ Per (B1 III).
The cluster members spread over about $15' \times 15'$ on the sky. 
The optically brightest and most massive cluster member
is the B5\,V star BD +31\degr\,643.

In a large number of observational studies, more than $300$ individual cluster members 
have been identified and well characterized so far.
The pioneering study of \cite{Herbig54.1} discovered 16 H$\alpha$ emitting
stars in IC~348. Later, \cite{Herbig98.1} identified about $100$ members by optical spectroscopy and
photometry as well as deep H$\alpha$ imaging.
The near-infrared photometric survey of 
\cite{Lada95.1}, and later, deeper infrared (IR) imaging studies were
used to construct the luminosity function of the cluster and provided
information on the stellar and sub-stellar mass function 
\citep{Lada98.1, Muench03.1, Preibisch03.1}.

By means of optical and IR spectroscopy and photometry, the investigations of 
\cite{Luhman98.1, Luhman99.1, Luhman03.2, Luhman05.3} progressively increased the completness of the
census of the young stellar population in IC\,348. These studies yielded a sample of $302$ well 
characterized members, for most of which individual stellar masses and ages can be reliably
estimated. 
Although the cluster contains a few embedded stars 
this extinction-limited ($A_{\rm V} \le 4$\,mag) sample of members  
is nearly complete 
down to masses of $M \ge 0.03\,M_\odot$;
it also contains $23$ spectroscopically identified young brown dwarfs (BDs), 
with spectral types as late as M9 and estimated masses down to $0.015\,M_\odot$.
Recently, \cite{Burgess09.1} discovered a young T-dwarf candidate
in IC\,348. If its status as a cluster member is confirmed, the
derived spectral type of $\sim$~T6 suggests a mass of only a few Jupiter masses.

IC\,348 was also extensively studied in the near- and mid-IR with the {\em Spitzer Space Telescope}.
\cite{Lada06.1} constructed optical/IR spectral energy distributions (SEDs) for all the cluster members
and investigated both the frequency and nature of the circumstellar disk population in the cluster.
\cite{Muench07.1} performed a {\em Spitzer} census of IC\,348 and found that about $50$\% of the known 
cluster members possess circumstellar disks.
\cite{Currie09.2} presented new, deep MIPS photometry of IC\,348
that allowed a detailed characterization of the circumstellar disks around the known cluster members.
Finally, millimetric observations directly revealed the presence and mass of $10$ disks around
IC\,348 members \citep{Lee11.1}. 
\cite{Dahm08.0} performed a spectroscopic investigation of accretion 
diagnostics for $40$ near solar-mass members of IC\,348
and derived accretion luminosities and rates for $14$ of these stars.

Based on all these studies, the basic properties of IC\,348 can be summarized as follows:
The mean age of the cluster members of $\sim 2-3$~Myr is very interesting because it corresponds 
to the time where the structure of the disks of most young stellar objects 
changes from primordial, rather massive accretion disks
to transitional and debris disks, and the point in time when
planetary formation is thought to occur.  
Furthermore, it offers interesting opportunities for comparisons to
younger clusters like the 
Orion Nebula Cluster \citep[$1-2$\,Myr; ][]{Hillenbrand97.0}
or the $\rho$~Oph cluster \citep[$\leq 1$\,Myr; ][]{Luhman99.2}, 
and older clusters and associations such as Upper Scorpius 
\citep[$\sim 5$\,Myr; ][]{de-Geus89.0}.

The extinction of the cluster members ranges from $A_V \sim 1\,{\rm mag}$ 
up to $A_V \ga 10$\,mag, with a mean value of $\sim 3.5$\,mag.
The majority of the cluster members is in the T Tauri stage of pre-main sequence evolution, 
and more than $70$ stars can be classified as ``classical T Tauri stars'' 
(CTTS) on the basis of their H$\alpha$ emission. In the central parts of the cluster the active star formation phase seems to be finished, 
and the cluster population represents the outcome of a recent star formation event.
About $10'$ to the south-west of the cluster center, however, a dense cloud core 
is found which contains several deeply embedded IR sources with
extinctions exceeding $\sim 20$ mag in $A_V$ as well as the very young molecular hydrogen jet HH~211 
\citep{McCaughrean94.1} and the IC~348~MMS outflow \citep{Eisloeffel03.1}. 
The star formation process is still underway in this region.

IC\,348 is also well studied in the X-ray regime. The first X-ray observations of IC\,348 by
\cite{Preibisch96.1} were performed with the
{\em ROSAT} observatory and led to the discovery of 116 X-ray sources.
In September 2000, a deep X-ray image of IC\,348 was obtained with the Advanced CCD Imaging 
Spectrometer (ACIS) on board the {\em Chandra X-Ray Observatory} and led to the detection of $215$ 
X-ray sources in the $17' \times 17'$ field-of-view.
These {\em Chandra} observations were analyzed by 
\citet[][PZ01 hereafter]{Preibisch01.1} and \citet[][PZ02 hereafter]{Preibisch02.1}. 
In February 2003, IC~348 was observed with {\em XMM-Newton}
\citep{Preibisch04.1}. 
The {\em XMM-Newton} observation was
strongly affected by solar particle flares
and consequently the sensitivity was significantly reduced. 
Therefore, only a small number of new X-ray sources were 
detected in the field previously observed with {\em Chandra}.
Two more recent {\em Chandra} observations centered on 
the south-western corner of IC\,348 were analysed by \cite{Forbrich11.1}.
They detected about one-third of the known cluster members in this area. 

While the analysis of these X-ray studies yielded numerous
interesting results about the X-ray properties of the young stars in IC\,348,
most of the above mentioned optical/IR studies of the cluster  
were not yet available at the time. 
(The recent {\em Chandra} 
study of the fields in the south-west of the cluster by \cite{Forbrich11.1} 
focused on a comparison of radio and X-ray emission of YSOs 
without detailed consideration of the stellar properties.)
The very substantial increase of optical/IR information about the cluster population 
during the last $\sim$ five years clearly warrants a repeated effort to study the relations between
X-ray properties and stellar parameters.
Another important aspect in this context is the availability
of three new deep {\em Chandra} observations of IC~348,
obtained in 2008 and spatially overlapping with the original field. 
The resulting {\em Chandra} `mosaic' 
comprises more than $90$\,\% of the cluster population; 
see Sect.~\ref{subsect:results_det} for more details. 

During the last few years, several very deep and extensive
X-ray studies of other low-mass star forming regions have been performed, which
provide a good basis for comparisons of the X-ray properties
of young stars at different ages and in different environments.
The first of these studies was the $Chandra$ Orion Ultradeep Project (COUP), 
a 10-day long observation of the Orion Nebula Cluster (ONC)  
with $Chandra$/ACIS \citep[see][]{Getman05.1}, the
deepest and longest X-ray observation ever made of a young stellar cluster.
A detailed 
investigation of the relation between the optical and X-ray properties 
of the $\sim 600$ X-ray detected optically visible and well characterized 
cluster members was presented in \citet{Preibisch05.1} 
and \citet{Stelzer05.1}.
The XMM-Newton Extended Survey of the Taurus Molecular Cloud (XEST)
covered the densest stellar populations in a 5 square degree region
of the Taurus Molecular Cloud \citep[see][]{Guedel07.3}
and provided X-ray data on 110 optically well characterized young stars.

The analysis presented in this paper uses all available {\em Chandra}
observations of IC\,348 to increase the spatial coverage of the cluster and the 
sensitivity in the region where the pointings overlap.
The X-ray source list is cross-correlated with an updated membership catalog that we have 
compiled from the literature. This yields stellar parameters, disk and accretion
indicators and rotation rates for the majority of X-ray emitting YSOs in IC\,348. 
We examine the dependence of X-ray emission level on these parameters 
and compare the results to X-ray studies of other nearby low-mass star forming regions
such as Orion and Taurus
with the scope to understand the origin of activity in pre-main sequence stars.

\section{The catalog of IC\,348 members}\label{sect:cat}

\subsection{The data base}\label{subsect:cat_data}

Since the studies by PZ01, 
the membership list of IC\,348 has changed significantly with more than $130$ 
new members confirmed by signatures of youth in optical low-resolution spectra and their
position in the HR diagram above the $10$\,Myr isochrone \citep[][henceforth LSM\,03]{Luhman03.2}. 
LSM\,03 considered membership complete in the central 
$16^\prime \times 14^\prime$ for $M > 0.03\,M_\odot$ and $A_{\rm V}< 4$\,mag. 
However, 
additional members have been identified on the basis of IR spectroscopy 
\citep[][henceforth LLM\,05]{Luhman05.7},
and {\em Spitzer} mid-IR photometry \citep{Muench07.1}, 
of which about a dozen in the LSM\,03 completeness area. 
Several objects that had late-M spectral types estimated previously from HST near-IR photometry 
\citep{Najita00.1}
have been rejected by LSM03 as cluster members on the basis of their optical spectra. 

{\em Spitzer} photometry from IRAC and MIPS for the members from LSM\,03 was 
discussed by \cite{Lada06.1}.
They examined the slope of the SED between $3.6...8\,\mu$m
for $263$ objects with IRAC photometry in all bands, and distinguished protostars, disk-bearing stars
and disk-less stars in four categories that we describe in Sect.~\ref{subsect:cat_derived}.

The present paper deals with the more than $90$\,\% of  
known IC\,348 members  that are within the field of the 
presently available {\em Chandra} observations. 
The catalog of optical/IR cluster members consists of the objects from Table~2 in LSM03, 
Table~2 in LLM05 and Table~1 in \cite{Muench07.1}.  
These catalogs provide spectral types, effective temperatures ($T_{\rm eff}$),
extinctions ($A_{\rm J}$ or $A_{\rm V}$), bolometric luminosities ($L_{\rm bol}$),
and optical/near-IR photometry. 
For the objects from Luhman's papers {\em Spitzer} photometry is extracted from \cite{Lada06.1}. 
Accretion rates are collected from \cite{Dahm08.0, Muzerolle03.1, Mohanty05.1},
and H$\alpha$ equivalent widths are those listed in LSM03.

\subsection{Derived properties}\label{subsect:cat_derived}

We have complemented the data base described above by several properties derived from
the literature data. 

First, YSOs are classified in two different ways, (i) on the basis of the IR slope of their SED and (ii)
on the basis of their H$\alpha$ equivalent width. 

For the SED classification the slopes $\alpha_{\rm 3-8\,\mu m}$ 
derived by \cite{Lada06.1} and \cite{Muench07.1} are used together with the boundaries defined by 
\cite{Lada06.1} that separate 
YSOs in different evolutionary phases 
(see Table~\ref{tab:yso_class}).
The four groups represent Class\,I objects (protostars), Class\,II (stars with
thick disks), Class\,II/III (stars with "anemic" disks), and Class\,III (disk-less stars).
The group termed "anemic" disks fills the gap between disk-bearing and disk-less stars
in {\em Spitzer}/IRAC color-color diagrams. This group shows a very small spread of [3.6]-[8.0] 
color which forms the lower envelope to thick disk stars. 
\begin{table}
\begin{center}
\caption{Definition of YSO class} 
\label{tab:yso_class}
\begin{tabular}{lr} \\ \hline
Class  & $\alpha_{\rm 3-8\,\mu m}$\tablefootmark{*} \\ \hline
I      & $>-0.5$ \\
II     & $-0.5....-1.8$ \\
II/III & $-1.8...-2.56$ \\
III    & $<-2.56$ \\
\hline
\end{tabular}
\tablefoot{\tablefoottext{*}{Slope of the SED from $3-8\mu$m (adopted from \protect\cite{Lada06.1} and \protect\cite{Muench07.1}).}}
\end{center}
\end{table}

\begin{figure}
\begin{center}
\resizebox{9.cm}{!}{\includegraphics{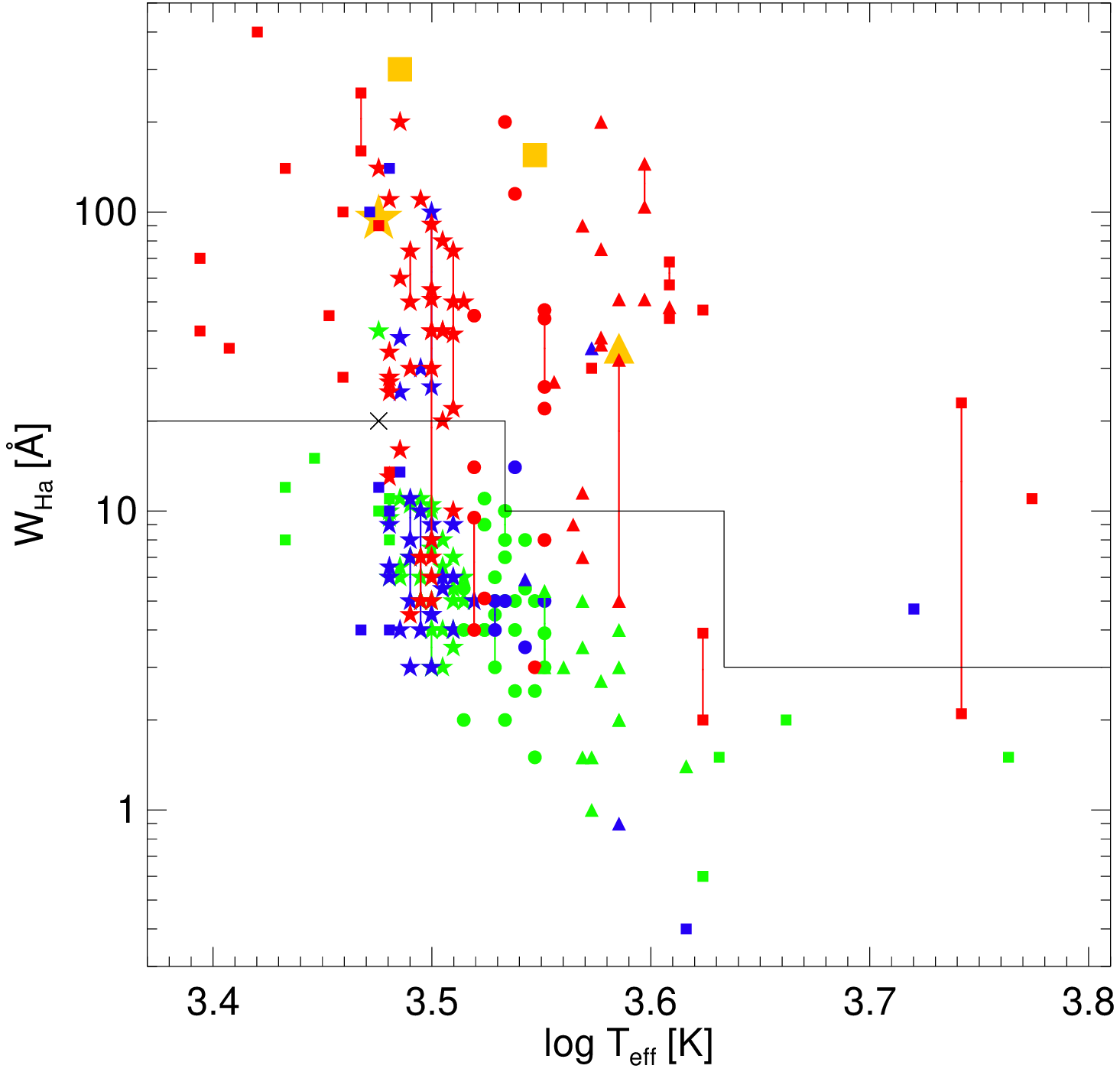}}
\caption{Equivalent width of H$\alpha$ vs. effective temperature. 
The color-codes refer to the {\em Spitzer} YSO classes: Class\,III (green; light grey in b/w print), Class\,II/III (blue; dark grey in b/w print),
Class\,II (red; medium grey in b/w print) and Class\,I (yellow and larger). Different mass ranges are symbolized with different
plotting symbols: $0.1-0.25\,M_\odot$ (star symbols), 
$0.25-0.6\,M_\odot$ (circles), $0.6-1.2\,M_\odot$ (triangles) and all others (squares). 
Vertical lines connect different epochs of $W_{\rm H\alpha}$ measurements for individual stars. 
The black line separates accreting
cTTS (top) from non-accreting wTTS (bottom). The cross on the black line denotes spectral type
M6. For objects with $T_{\rm eff}$ smaller than this value \protect\cite{White03.1} have
placed the dividing line between cTTS and wTTS at $40$\,\AA. However, our boundary of $20$\,\AA~
is in better agreement with the separation of Class\,II and Class\,III sources.} 
\label{fig:wha_teff}
\end{center}
\end{figure}

The distinction of classical and weak-line T Tauri stars (henceforth cTTS and wTTS)
on the basis of the strength of H$\alpha$ emission must consider the fact that 
the H$\alpha$ equivalent width ($W_{\rm H\alpha}$) is not equivalent to the line flux  
as a result of its dependence on the continuum, which declines towards later spectral types. 
Therefore,
when using $W_{\rm H\alpha}$ as an accretion indicator the threshold between cTTS and wTTS must be 
set at successively higher values for stars of later spectral types.
We use thresholds of $3$, $10$, and $20$\,\AA~ for stars earlier than K5, between K6 and M3, and later than M3,
respectively. This is similar to the criteria introduced by \cite{White03.1}, who included an additional group 
with $W_{\rm H\alpha}>40$\,\AA~ for stars later than M6. 
The omission of this last step is justified
by the separation of Class\,II and Class\,III sources in Fig.~\ref{fig:wha_teff}. 

A comparison of the two ways used to classify YSO evolution gives clues to the
connection between the dissipation of dust and gas during the pre-main sequence. 
\cite{Lada06.1} found that in IC\,348 there is a close relation between 
H$\alpha$ equivalent width and IR excess. 
We confirm this in Fig.~\ref{fig:wha_teff} using a more sophisticated division between
accretors and non-accretors that depends on spectral type. 
The majority of Class\,II sources are classified as 
accretors on the basis of their H$\alpha$ emission. 
Only one Class\,III source has an H$\alpha$ equivalent width in the range typical
for accretors, and only $6$ out of $28$ Class\,II/III sources. 
However, next to the majority of Class\,II/III sources, 
several Class\,II sources 
are classified as `weak-lined' stars according to our criterion, i.e. they have disks
but are not considered to be actively accreting any more presumably 
because the gas has dispersed. 
These objects are likely the more evolved among the Class\,II and Class\,II/III stars. 
Another possible explanation for the presence of 
Class\,II sources below the cTTS cutoff line is variable
accretion that would allow for some disk-bearing Class\,II and Class\,II/III objects 
to appear temporarily as wTTS. Stars with more than one H$\alpha$ measurement 
(connected by vertical lines in Fig.~\ref{fig:wha_teff}) 
show indeed, that H$\alpha$ emission is strongly variable, and the equivalent widths measured
at different epochs for some Class\,II objects do cross the boundary between cTTS and wTTS. 
For those stars we have used the highest value of $W_{\rm H\alpha}$ 
as criterion for their cTTS/wTTS classification. 


It can also be seen from Fig.~\ref{fig:wha_teff} that a few 
sources classified as Class\,I have not only a determined spectral 
type but also detected H$\alpha$ emission,  
both unexpected for true protostars that ought to be invisible in the 
optical due to high extinction of their envelopes. 
In total, among the $15$ Class\,I sources known in IC\,348, 
a spectral type has been detected for ten of them 
and five have a measurement for H$\alpha$ emission. The H$\alpha$ equivalent
widths of these stars are all above our `accretor threshold'.
These objects are probably no genuine protostars but
T\,Tauri stars with massive disks 
being responsible for their shallow slope in the mid-IR. 
In fact, \cite{Muench07.1} state that the distinction between highly
flared Class\,II disks and disk/envelope systems may require observations
at wavelengths longward of $10\,\mu$m. Such measurements are not available
for the majority of our sample stars, and in any case they are not taken into
account in the definition of YSO classes as described in 
Table~\ref{tab:yso_class}.

\begin{figure*}
\begin{center}
\includegraphics[width=18cm]{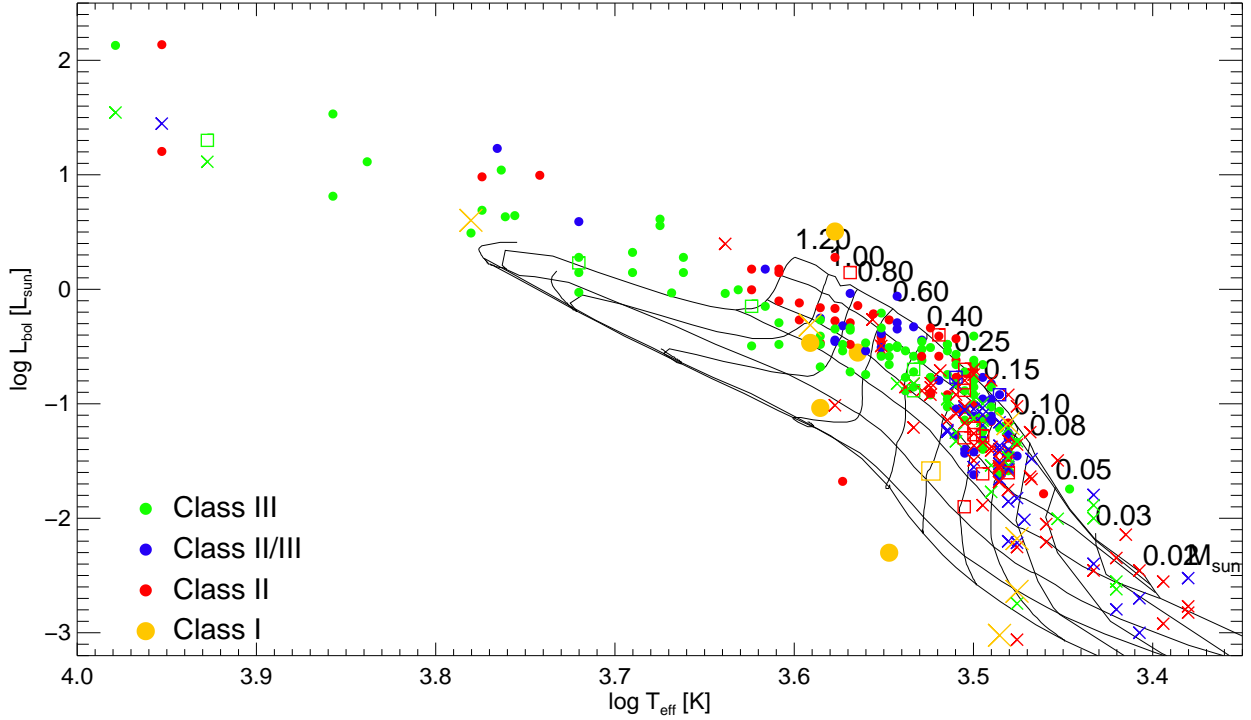}
\caption{HR diagram for IC\,348 compared to evolutionary tracks of
\protect\cite{Baraffe98.1} and \protect\cite{Chabrier00.2}. Isochrones are shown for
$1$, $2$, $5$, $10$, $30$, $100$, $1000$\,Myr. 
The tracks are labeled in units of solar mass. 
X-ray detections are represented by filled circles and non-detections as crosses. 
IC\,348 members outside the {\em Chandra} fields are shown as open squares. 
The objects are color-coded according to their YSO class 
(see definition in Sect.~\ref{subsect:cat_derived}).}  
\label{fig:hrd}
\end{center}
\end{figure*}

We have seeked to determine masses for all stars with tabulated 
$L_{\rm bol}$ and $T_{\rm eff}$ 
from an interpolation of evolutionary models. Different sets of models are known
to be inconsistent between each other. 
LSM03 have shown that the \cite{Baraffe98.1} and \cite{Chabrier00.2} models
yield the best description for the low-mass stars in IC\,348. 
These are also the only models that include very low mass stars ($< 0.1\,M_\odot$)
and extend into the BD regime. 
The HR diagram for IC\,348 is shown in Fig.~\ref{fig:hrd} where we anticipate 
the result from the cross-correlation with X-ray sources (see Sect.~\ref{sect:xrays}). 
Our choice of models goes at the expense
of an analysis of stars with $M > 1.2\,M_\odot$. However, as can be seen from
Fig.~\ref{fig:hrd} the vast majority of IC\,348 members have lower mass. 
No mass could be determined for a few stars below the zero-age main-sequence.
These are probably stars with edge-on disks seen in scattered light, consistent with their
classification as YSOs with circumstellar matter (either Class\,II or Class\,I). 
%

\section{X-ray data analysis}\label{sect:xrays}

\begin{figure*}
\begin{center}
\includegraphics[width=16cm]{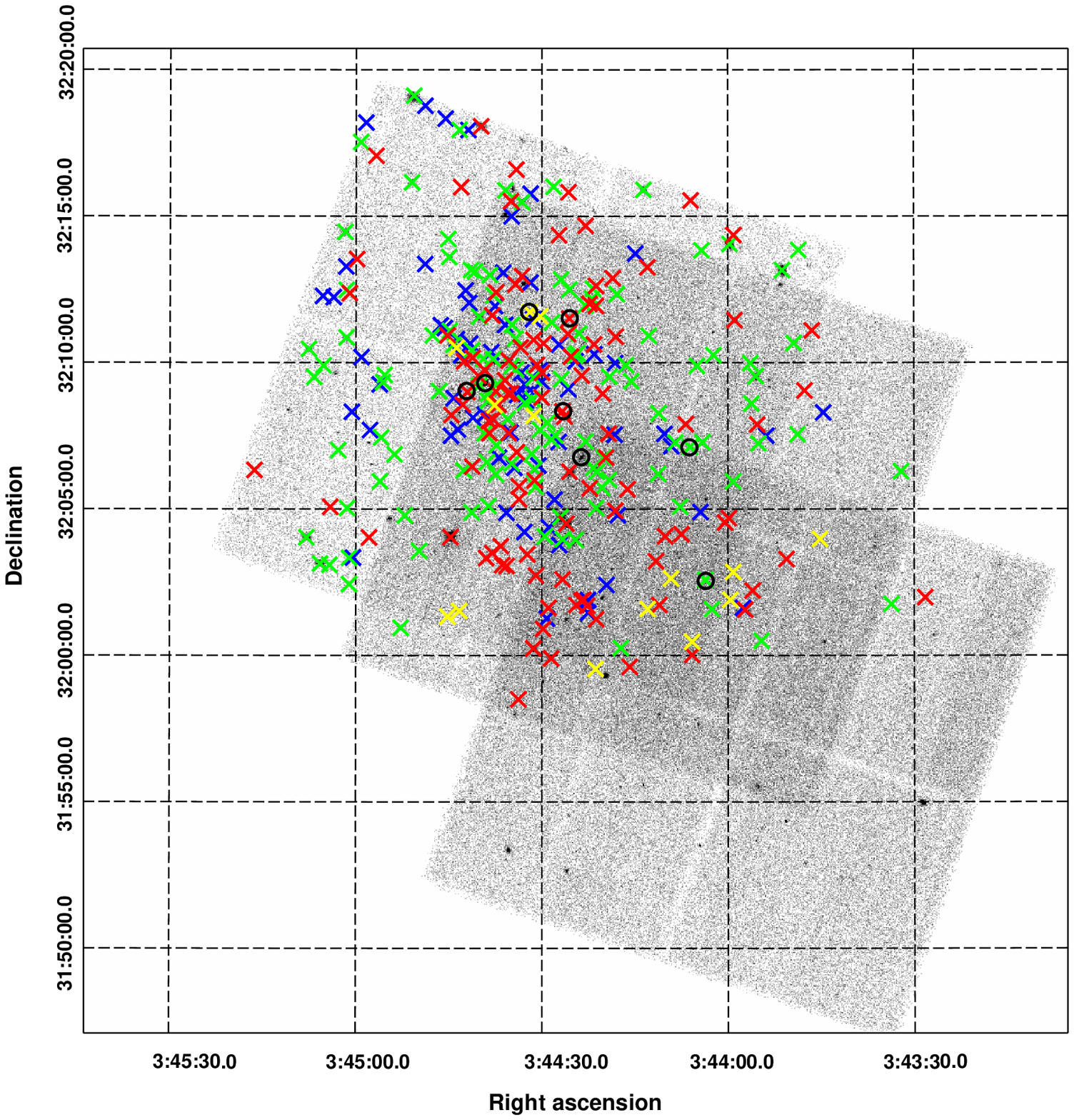}
\caption{Merged {\em Chandra}/ACIS-I image created from the four observations listed in 
Table~\ref{tab:observations}. IC\,348 members in different YSO phases are color-coded as
in Fig.~\ref{fig:wha_teff}. Cluster members without known YSO class are shown as black
circles; many of these are secondary components in binaries.}
\label{fig:acis_fov}
\end{center}
\end{figure*}

IC\,348 was observed in September 2000 (ObsID 606, PI: Th.~Preibisch) and three times in March 2008 
(ObsIDs 8584 PI: N.~Calvet, 8933 \& 8944 PI: S.~Wolk) with the Imaging Array of the 
\textit{Chandra} Advanced CCD Imaging Spectrometer (ACIS-I), that provides a field of view of 
$17' \times 17'$ on the sky. 
The aimpoint of observation 
%
8584 was offset to the southwest with respect to pointing 606 and  
%
observations 8933 and 8944 were offset even more to the southwest; see Fig.~\ref{fig:acis_fov} 
for a graphical representation of the Chandra observations. 
%
The total net exposure time of all four observations was 182\,860~s (50.79 hours). 
Table~\ref{tab:observations} contains detailed information on the 4 observations. 
\begin{table*}
\centering                          
\caption{{\em Chandra} observation log}             
\label{tab:observations}      
\begin{tabular}{c c c c c c c c c}        
\hline\hline                 
Obs.Id. & \multicolumn{2}{c}{Date Start [UT]} & \multicolumn{2}{c}{Date End [UT]} & Exposure time & Level 2 events & $\alpha_{\rm 2000}$\,[h:m:s] & $\delta_{\rm 2000}$\,[$^\circ$:$^\prime$:$^{\prime\prime}$] \\    
\hline                        
   606  & 2000-09-21 &19:58:42 & 2009-09-22 & 11:09:29& 52\,960 s&  347\,817 & 03:44:30.00 & +32:08:00.0 \\	 
   8584 & 2008-03-15 &09:02:20 & 2008-03-15 & 23:33:38& 50\,140 s&  582\,252 & 03:44:13.20 & +31:06:00.0 \\	 
   8933 & 2008-03-18 &17:35:07 & 2008-03-19 & 15:12:52& 40\,120 s&  294\,144 & 03:43:59.90 & +31:58:21.7 \\	 
   8944 & 2008-03-13 &17:53:02 & 2008-03-14 & 05:25:12& 38\,640 s&  300\,328 & 03:43:59.90 & +31:58:21.7 \\	 
   
\hline                                   
\end{tabular}
\end{table*}
The roll angles (i.e.~the orientation of the detector on the sky) are equal ($288-289^\circ$) for the three observations from 2008. For observation 606 the roll angle was $109.6^\circ$ and hence the detector was rotated $180^\circ$. Therefore sources are only at the same detector position for observations 8933 and 8944. Two of the CCDs of the ACIS-S spectroscopic array were also turned on during the observations. However, since the PSF at the large off-axis angles at these detectors is strongly degraded, their point-source sensitivity is reduced; analysis was performed without using the data from ACIS-S. 
The basic data products of our observation are the four Level~2 processed event lists provided by the pipeline processing at the \textit{Chandra} X-ray Center, that list the arrival time, location on the detector and energy for each of the 1\,524\,541 detected X-ray photons. 
We combined the four pointings with the {\it merge\_all} script, a {\em Chandra} contributed software that makes use of standard {\it CIAO}\footnote{Chandra Interactive Analysis of Observations, version 4.2:\\ http://cxc.harvard.edu/ciao/index.html} tools.
The merged events file is shown in Fig.~\ref{fig:acis_fov}. Overlaid are crosses at 
the positions of the IC\,348 members showing the spatial distribution of the YSOs.

\subsection{Source detection and X-ray source catalog}\label{subsect:xrays_detect}

The source detection was carried out in a two-step process. The first detection step was performed in a rather aggressive manner in order to find even the weakest possible sources, deliberately accepting some degree of false detections. In the second step, this list of potential sources was then cleaned from spurious detections by a detailed individual analysis. We employed the {\sc wavdetect} algorithm \cite[a {\it CIAO} mexican-hat wavelet source detection tool; ][]{Freeman02.1} for locating X-ray sources in our merged image, and used a detection threshold of 10$^{-5}$ and wavelet scales between 1 and 16 pixels. The result was a list of 372 prospective sources. 
Three sources were added, which have been detected with {\sc wavdetect} in an \textit{XMM-Newton} observation of IC~348. 
Another 17 possible sources were added, which were identified by cross-correlating \textit{UKIRT} observations with the \textit{Chandra} image by eye. 
These $20$ sources were first added as potential sources to the sourcelist for ACIS
Extract. ACIS Extract then constructs an extraction region at the position of
these sources and estimates the probability for being a source. 
The result was a final catalog with 393 potential X-ray sources. To clean this catalog from spurious sources, we performed a detailed analysis of each individual candidate source with the ACIS Extract (AE hereafter) software package\footnote{http://www.astro.psu.edu/xray/docs/TARA/ae\_users\_guide.html} \citep{Broos10.1}. A full description of the procedures used in AE can be found in \citet{Getman05.1}, \citet{Townsley03.1}, and \citet{Broos07.1}.
The following three steps were performed by AE in order to prune our input catalog from spurious detections (including afterglows):
\begin{enumerate}
\item{Extraction regions were defined such as to include 90\% of the 
source photons in the PSF 
(or less in the case of other nearby sources), 
and source events were extracted from this region.
ARFs and RMFs are constructed for each source, and the ARFs are corrected for the light missed 
by the finite extraction regions.} 
\item{Local background events were extracted after masking all the sources in the catalog.}
\item{The Poisson probability ($P_{B}$) associated with the ``null hypothesis'', i.e.~that no source exists and the extracted events are solely due to Poisson fluctuations in the local background, was computed for each source.}
\end{enumerate} 

After 6 iterations of this pruning procedure our final catalog consisted of 290 X-ray sources, for which the probability of being a background fluctuation is less than $0.1$\,\%. For 287 sources of those the probability is less than $0.01$\,\% and 266 sorces have a very low probability of $10^{-5}$.
Some of the basic source properties derived by the AE software, 
such as the net (i.e., background-subtracted) counts in various energy bands, the median photon energy ($E_{\rm med}$), statistical test for variability, and a measure of the incident photon flux,
are reported in Table~\ref{tab:src_properties_main}. 
Sources are sorted by increasing right ascension and identified by their sequence number (Col.~1) or their IAU designation (Col.~2).

\subsection{Calculation of X-ray fluxes}\label{subsect:xrays_flux}

An accurate determination of the intrinsic X-ray source luminosities requires good knowledge of the X-ray spectrum. However,
for the majority of the X-ray sources the number of detected photons is too low (less than 100 net counts) for a detailed spectral analysis. 

An estimate of the intrinsic, i.e.~extinction corrected, X-ray luminosity for 
sources that are too weak for a detailed spectral analysis can be obtained 
with the \textit{XPHOT} software\footnote{http://www.astro.psu.edu/users/gkosta/XPHOT/}, developed by \citet{Getman10.1}. \textit{XPHOT} is based on a 
non-parametric method for the calculation of fluxes and absorbing X-ray 
column densities of weak X-ray sources. 
Our procedure is based on a grid of simulated X-ray spectra 
with the following parameters: two temperatures, the emission measure ratio of
the two spectral components, and the absorbing column.
The X-ray extinction and intrinsic flux of a given X-ray source 
are estimated from a sophisticated comparison scheme 
between the observed median energy 
and the observed source flux and analogous values obtained from 
the simulated spectra. 
This method requires 
at least 4 net counts per source (in order to determine a meaningful value 
for the median energy) and can thus be applied to 250 of our 290 sources. 

An estimate of the \textit{observed} (i.e.~\textit{not} the intrinsic) X-ray flux is also computed by AE. This quantity, called $FLUX2$ and given in units of ${\rm photons/s/cm^2}$, is calculated from the number of detected photons and using a mean value of the instrumental effective area (through the Ancillary Response Function, ARF) over energy. 
It should be noted that the $FLUX2$ values suffer from a systematic uncertainty with respect to the true incident flux, because the use of a mean ARF is only correct in the hypothetical case of a flat incident spectrum, an assumption that is probably not fulfilled. Nevertheless, the $FLUX2$ represents the best estimate for the photon flux 
that can be obtained for comparatively weak sources. 
The product of $FLUX2$ and $E_{\rm med}$ provides an estimate for the energy flux of the source. 
When compared to the fluxes derived with XPHOT a rather good linear correlation is found 
but the product $FLUX2 \cdot E_{\rm med}$ results in systematically higher values. 
We calibrated the $FLUX2$ values using the correlation with the XPHOT fluxes 
to obtain source fluxes for the $40$ sources that are too faint for XPHOT.  

This way, the X-ray flux of all $290$ detected sources has been determined in a consistent way. 
We cross-correlate the X-ray source list with the catalog of IC\,348 members from
Sect.~\ref{sect:cat} using a match radius of $1.5^{\prime\prime}$. 
Larger cross-correlation radii did not substantially increase the number of matches, 
e.g. a radius of $3.0^{\prime\prime}$ yields only four additional identifications, 
an increase of the X-ray sample for IC\,348 members 
of about $2$\,\%. We are, therefore, confident that nearly
all X-ray emitting IC\,348 members have been recognized by our procedure. 
The flux of all X-ray detected IC\,348 members is given in Table~\ref{tab:fluxes}. 
A distance of $310$\,pc \citep{Herbig98.1} is assumed for conversion of the flux to luminosity. 
The resulting intrinsic X-ray luminosities range from $10^{28.38}$ to $10^{31.93}$~erg~s$^{-1}$.
%
\begin{table}\begin{center}
\caption{X-ray fluxes and upper limits}
\label{tab:fluxes}
\begin{tabular}{ccrr} \hline
{\em Chandra} ID & ID\tablefootmark{a} & \multicolumn{1}{c}{$\log{f_{\rm x}}$} & flag\tablefootmark{b} \\
           &            & \multicolumn{1}{c}{[${\rm erg/cm^2/s}$]}  & \\ \hline
 &            1 & $<-14.34$ &  \\
034435.35+321004.5 &            2 & $-12.32$ &  \\
034450.64+321906.5 &            3 & $-12.10$ &  \\
034431.21+320622.1 &            4 & $-13.71$ &  \\
034426.03+320430.4 &            5 & $-12.34$ &  \\
034436.93+320645.4 &            6 & $-12.09$ &  \\
 &            7 & $<-14.35$ &  \\
 &            8 & $<-14.07$ &  \\
034439.15+320918.1 &            9 & $-12.25$ &  \\
034424.66+321014.8 &           10 & $-14.28$ &  \\
034507.96+320401.8 &           11 & $-12.32$ &  \\
 &           12 & $<-14.78$ &  \\
 &           12 & $<-14.63$ &  \\
034359.67+320154.1 &           13 & $-12.53$ &  \\
\hline
\end{tabular}
\tablefoot{
\tablefoottext{a}{Identifier from LSM03, LLM05 or \protect\cite{Muench07.1} for all IC\,348 members observed with {\em Chandra}.}
\tablefoottext{b}{Flag is set if the flux was obtained from AE ($FLUX2$).}
}
\end{center}\end{table}

\subsection{Determination of upper limits}\label{subsect:xrays_uls}

For those $129$ stars from LSM03, LLM05 and \cite{Muench07.1}, 
which had no counterpart in our X-ray source catalog, upper limits had to be determined. We did this using again the AE software package. Extraction regions were defined the same way as the 90\% contours of the local PSF to determine the number of counts in the target aperture and an estimate of the local background. To calculate upper limits for the received source counts from AE we used the program \textit{bayes.f} \citep{Kraft91.1}. This program performs the calculation of the upper limit on a Poisson process with background, incorporating the effects of uncertainty on the background and signal acceptance. Upper limits for the number of source counts were determined for a confidence level of 0.9. For each object we divided the upper limit for the source counts by the local exposure time. To obtain upper limits for the luminosity we determined a conversion factor 
($CF$) between count rate and luminosity. As countrate and XPHOT luminosity are 
correlated for T Tauri stars, this was done with a sample of 180 stars with known spectral 
type from our AE source catalog. 
The mean of the $CF$ is $3.74462\cdot10^{32}\,{\rm erg/counts}$.
The $CF$ depends essentially on absorption, and we verified that the sample of
detected stars from which it was calculated  has a similar mean $A_{\rm J}$ value as the sample
of undetected stars to which the $CF$ was applied. 
The upper limit fluxes for those latter ones 
are listed in Table~\ref{tab:fluxes} together with the fluxes of the detected IC\,348 members.

\section{Results}\label{sect:results}

\subsection{X-ray detection statistics}\label{subsect:results_det}

Our combined membership list comprises $345$ YSOs from spectral type B5 to M9.
Only $29$ of them are located outside the sky region observed with {\em Chandra}. 
We have X-ray detected $187$ YSOs ($59$\,\% of the observed sample);
see Table~\ref{tab:yso_statistics}. 
Many of the new members identified after the study of PZ01 are too faint for 
detection in the existing {\em Chandra} data, even after combining the original image
(Obs-ID\,606) with the more recent exposures. 

\begin{figure*}
\begin{center}
\parbox{18cm}{
\parbox{9cm}{
\includegraphics[width=9cm]{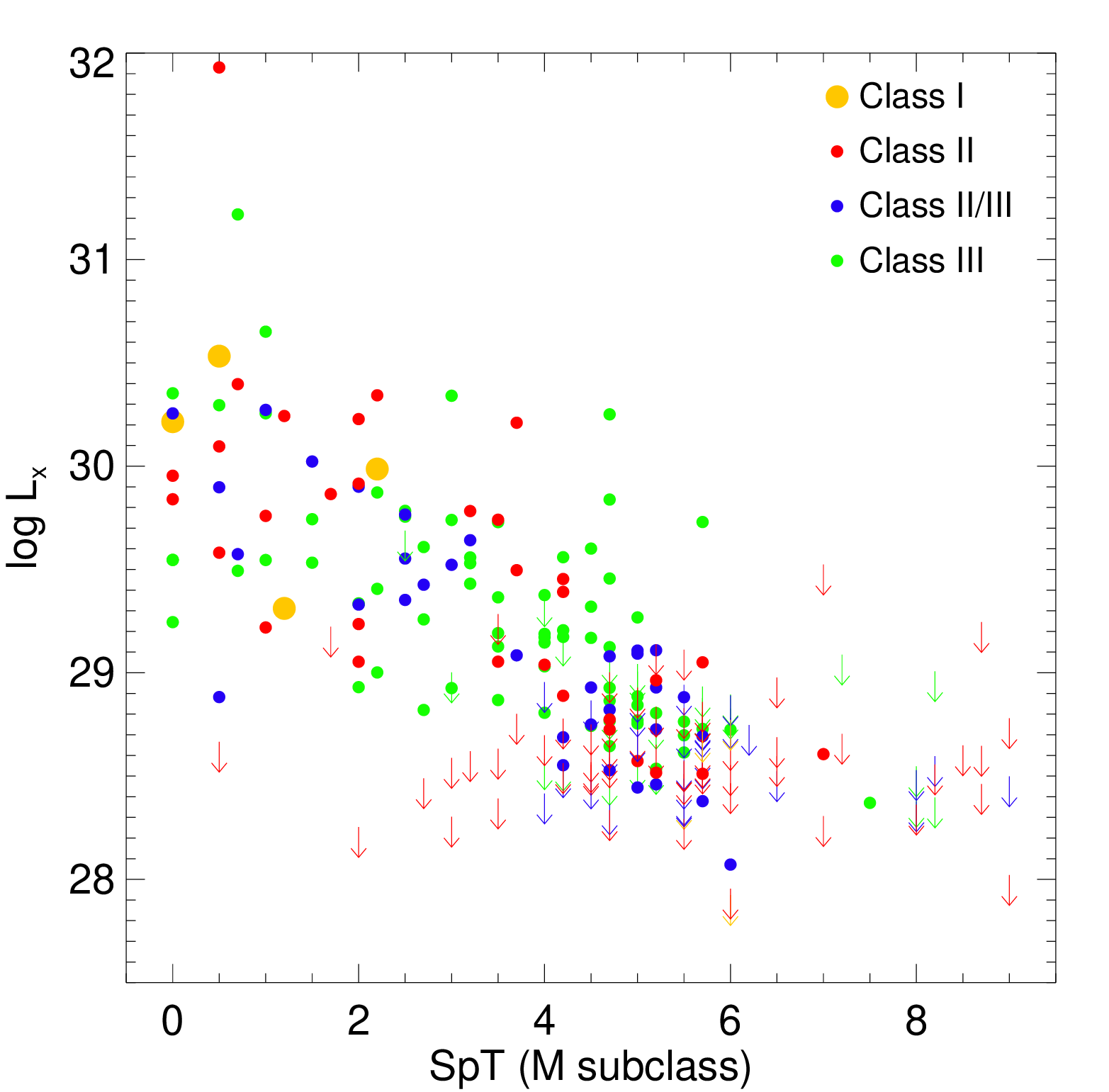}
}
\parbox{9cm}{
\includegraphics[width=9cm]{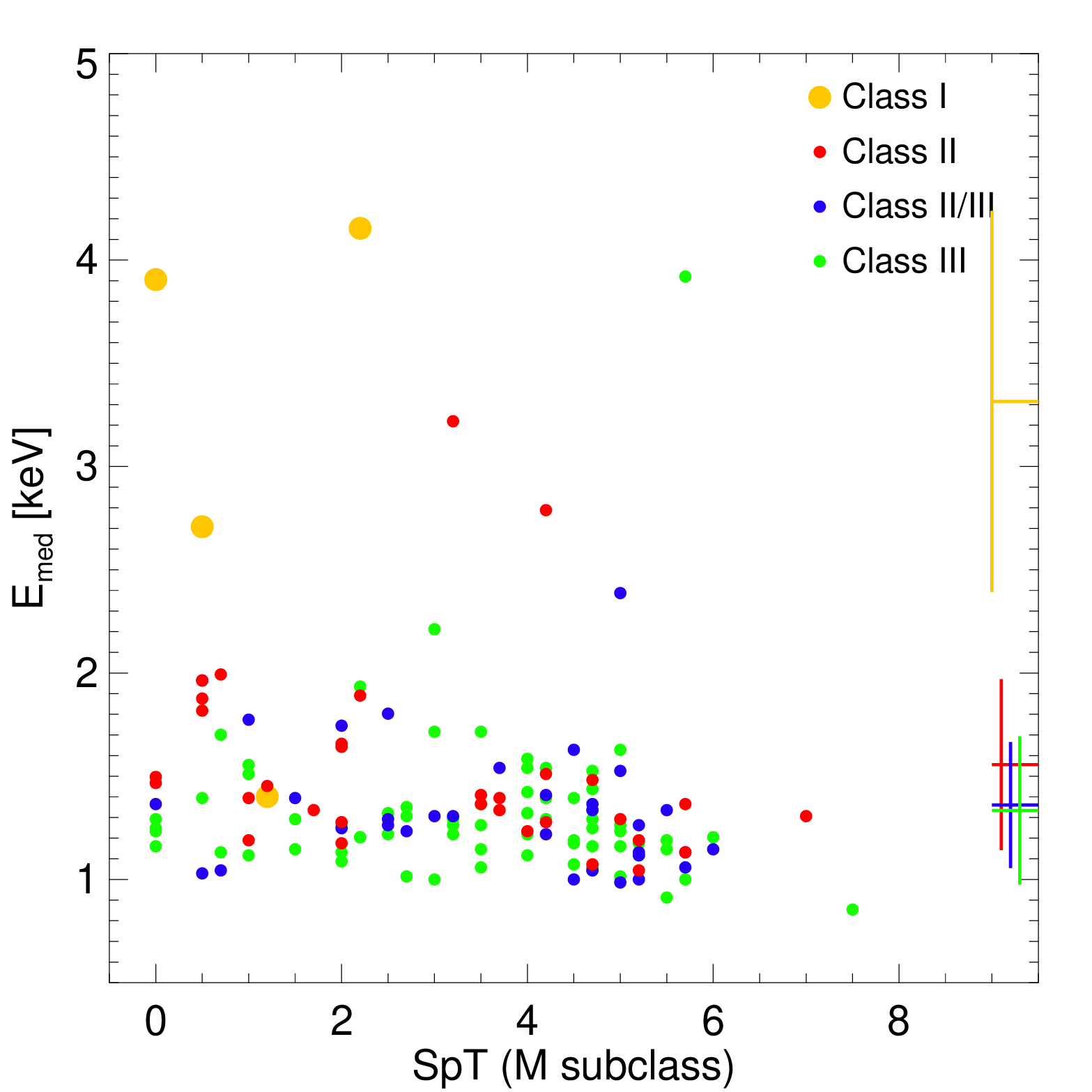}
}
}
\caption{X-ray luminosity and median energy for the low-mass end of IC\,348 versus spectral type.
The color codes refer to different YSO classes (see Fig.~\ref{fig:wha_teff}). The colored
lines in the right panel represent the mean of the median energy for each YSO group
irrespective of spectral type 
and its standard deviation, 
i.e. including also higher-mass stars for which individual data points are outside the plotted range.}
\label{fig:lx_mede_spt}
\end{center}
\end{figure*}

It is evident from the position of the $129$ X-ray undetected IC\,348 members in the HR diagram 
(Fig.~\ref{fig:hrd}) that these are mostly very-low mass stars and BDs. 
A closer look at the X-ray properties of the lowest mass objects in IC\,348 
is shown in Fig.~\ref{fig:lx_mede_spt} (left) 
where we display X-ray luminosities for spectral class M. 
Only $4$ of the $39$ known presumable BDs (defined here as objects with 
spectral type M6 and later) within the {\em Chandra} FOV are detected. 
The coolest X-ray detected IC\,348 member has spectral type M7.5. 
PZ01 claimed the X-ray detection of $7$ BDs and BD candidates. 
However, only two of them
were detected by their automatic source detection routine and the other five
by collecting the counts within a $3^{\prime\prime}$ region around the
optical position. 
In our new analysis we have used a more sophisticated source detection procedure. 
Moreover, the updates
of the stellar parameters in the recent IC\,348 membership studies have shifted two of 
the objects previously considered 
BDs to earlier spectral type, i.e. stellar mass (see LSM\,03).
%
\begin{table}
\begin{center}
\caption{X-ray detection statistics for different YSO classes in IC\,348}
\label{tab:yso_statistics} 
\begin{tabular}{llccccc}\hline
Mass Range [$M_\odot$] &                & I     & II     & II/III  & III & All \\ \hline
\multicolumn{1}{l}{all IC\,348 members} & & & & & \\ 
&$N_{\rm det}$   & 8     & 44     & 36      & 94  & 187 \\
&$N_{\rm ul}$    & 6     & 61     & 32      & 27  & 129 \\ 
&$\frac{N_{\rm det}}{N}$ \myrule & $0.57$& $0.42$ & $0.53$ & $0.78$ & $0.59$ \\ \hline
\multicolumn{1}{l}{$M < 0.1$} & & & & & \\ 
&$N_{\rm det}$   & 0     &  1     &  1      &  3  &  5 \\
&$N_{\rm ul}$    & 1     & 19     & 14      &  7  & 41 \\ 
&$\frac{N_{\rm det}}{N}$ \myrule & $-$   & $0.05$ & $0.07$ & $0.30$ & $0.11$ \\ \hline
\multicolumn{1}{l}{$0.1\,\leq M < 0.25$} & & & & & \\ 
&$N_{\rm det}$   & 0     & 10     & 17      & 23  & 50 \\
&$N_{\rm ul}$    & 2     & 28     & 15      & 13  & 58 \\ 
&$\frac{N_{\rm det}}{N}$ \myrule & $-$&  $0.26$ & $0.53$ & $0.64$ & $0.46$ \\ \hline
\multicolumn{1}{l}{$0.25\,\leq M < 0.6$} & & & & & \\ 
&$N_{\rm det}$   & 0     & 11     &  9      & 30  & 51 \\
&$N_{\rm ul}$    & 0     &  9     &  1      &  4  & 15 \\ 
&$\frac{N_{\rm det}}{N}$ \myrule & $-$&  $0.55$ & $0.90$ & $0.88$ & $0.77$ \\ \hline
\multicolumn{1}{l}{$0.6\,\leq M < 1.2$} & & & & & \\ 
&$N_{\rm det}$   & 4     & 14     &  6      & 21  & 45 \\
&$N_{\rm ul}$    & 1     &  2     &  0      &  0  &  3 \\ 
&$\frac{N_{\rm det}}{N}$ \myrule & $0.80$& $0.88$ & $1.00$ & $1.00$ & $0.94$ \\ \hline
\end{tabular} 
\end{center}
\end{table}

\subsection{X-rays and evolutionary stage}\label{subsect:results_yso}

\begin{figure*}
\begin{center}
\parbox{18cm}{
\parbox{9cm}{
\epsfig{file=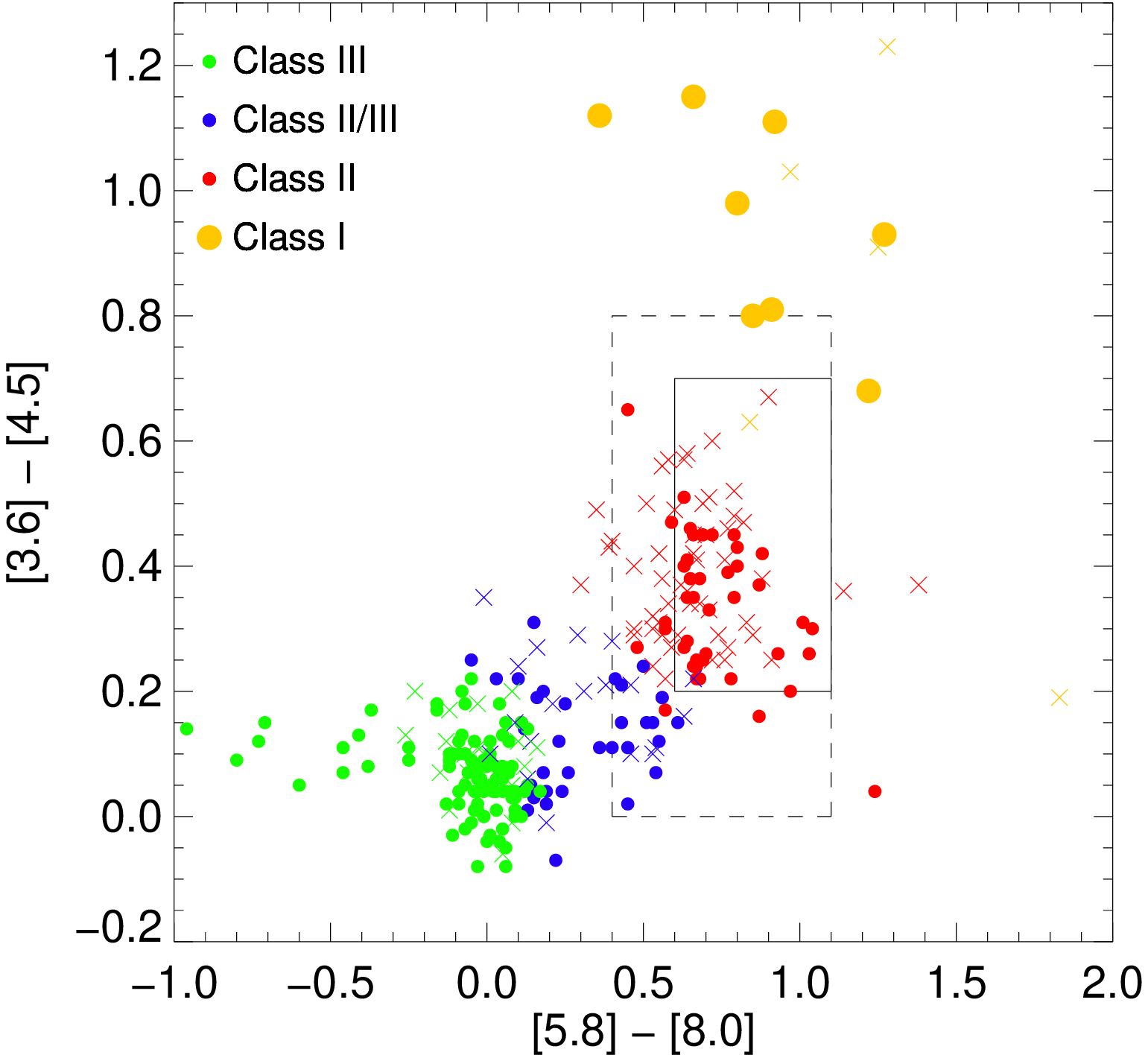,width=8.5cm}
}
\parbox{9cm}{
\epsfig{file=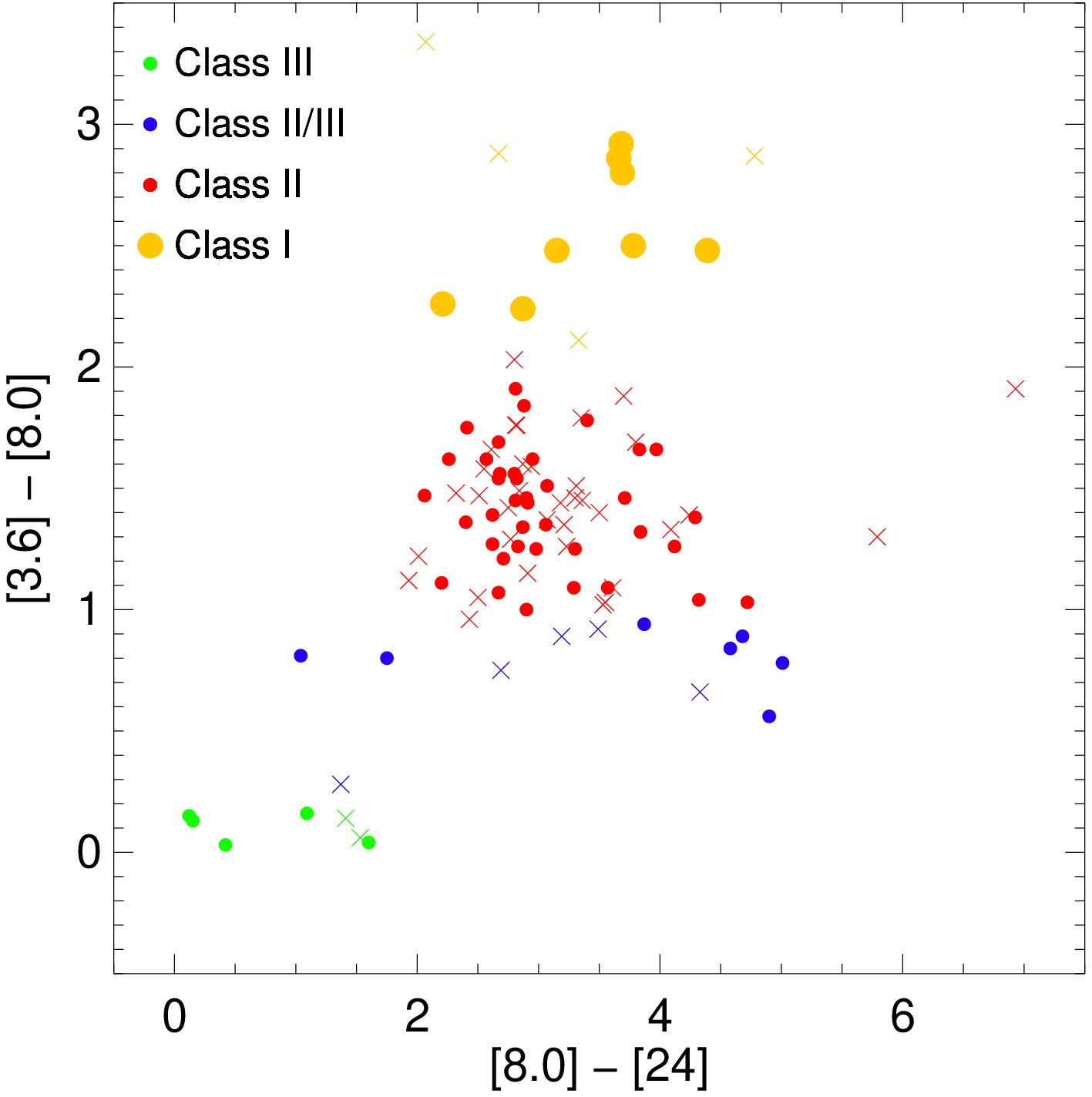,width=8.5cm}
}
}
\caption{{\em Spitzer} color-color diagrams for IC\,348 members in the {\em Chandra} fields. 
X-ray detections are represented by filled circles and non-detections as x-points. 
The color codes for different SED YSO classes 
are the same as in the previous figures and, for convenience, 
defined again in the top left of each panel. 
The boxes in the left figure show the areas defining Class\,II according to \protect\cite{Megeath04.1} and
\protect\cite{Hartmann05.1}; dashed and solid lines, respectively.}
\label{fig:spitzer_ccd}
\end{center}
\end{figure*}

Fig.~\ref{fig:spitzer_ccd} shows color-color diagrams for {\em Spitzer}/IRAC
and {\em Spitzer}/MIPS data 
for Lada Class\,I, II, II/III, and III sources. 
X-ray detected and undetected stars are represented by filled and open symbols respectively. 
In the left panel we overplot the Class\,II areas  
defined by Megeath et al. (2004) and by \cite{Hartmann05.1}.
As was shown already by \cite{Lada06.1},  
the YSO classification based on the SED slope corresponds to distinct 
regions in the {\em Spitzer}
color-color diagrams, in particular those involving [3.6] - [8.0]. 
This classification is in good agreement with the 
color-color selection for Class\,II sources introduced by Megeath et al. (2004), 
although the latter one includes a significant number of stars in the Class\,II area 
that are considered 
transition disks according to \cite{Lada06.1}. The Class\,II
criteria introduced by \cite{Hartmann05.1} are more conservative and include almost exclusively objects
with Lada Class\,II flag but they miss a good fraction of them. 

The X-ray detection statistics for the different YSO classes in IC\,348 
are summarized in Table~\ref{tab:yso_statistics}. 
Columns `$N_{\rm det}$' and `$N_{\rm ul}$' are the number of stars detected 
and undetected (`upper limits') with {\em Chandra}, respectively. 
In the last column we 
give the total number of stars, regardless of 
YSO class. Only $8$ confirmed IC\,348 members within the area surveyed by 
{\em Chandra} have no YSO classification. 
The detection fraction is highest for the most evolved stars, Class\,III,
regardless of mass. 
For each YSO class there is a clear trend of increasing detection fraction with increasing mass. 
In particular, for stars with $M > 0.6\,M_\odot$ all Class\,II/III and III sources are detected, and 
the total detection statistic regardless of YSO class is $> 94$\,\%. 
On the other hand only $\sim 10$\,\% of very low-mass 
objects with $M < 0.1\,M_\odot$ are X-ray detected. 

LSM\,03 consider membership complete in the central 
$16^\prime \times 14^\prime$ for $M > 0.03\,M_\odot$ and $A_{\rm V}< 4$\,mag,
but this estimate might be biased at the low-mass end against disk-less or strongly absorbed members  
as corroborated by the identification of further cluster members with IR spectroscopy and {\em Spitzer}
mid-IR photometry, several of which in the completeness area of LSM\,03 
(LLM05; \cite{Muench07.1}). 
The {\em Spitzer} color selection excludes by definition the identification
of Class\,III sources. 

The X-ray luminosities of the different YSO classes are compared in Sect.~\ref{subsect:results_xlf}.
Here we point at $E_{\rm med}$ as a rough measure for spectral hardness and/or extinction.
The median energy is displayed for X-ray detected M stars of IC\,348 in the right panel of 
Fig.~\ref{fig:lx_mede_spt}. For clarity only the spectral class M is displayed. The mean
of the median energy ($\langle E_{\rm med} \rangle$) 
for the full sample irrespective of spectral type 
is shown for each YSO class 
at the right end of the graph together with its standard deviation. 
It strikes that $\langle E_{\rm med} \rangle$ of Class\,I objects
is much higher than that of all other YSO classes. This is likely due to high
extinction from protostellar envelopes or thick disks absorbing the soft photons.

\subsection{X-ray and bolometric luminosity}\label{subsect:results_lbol}

\begin{figure}[t]
\begin{center}
\includegraphics[width=9cm]{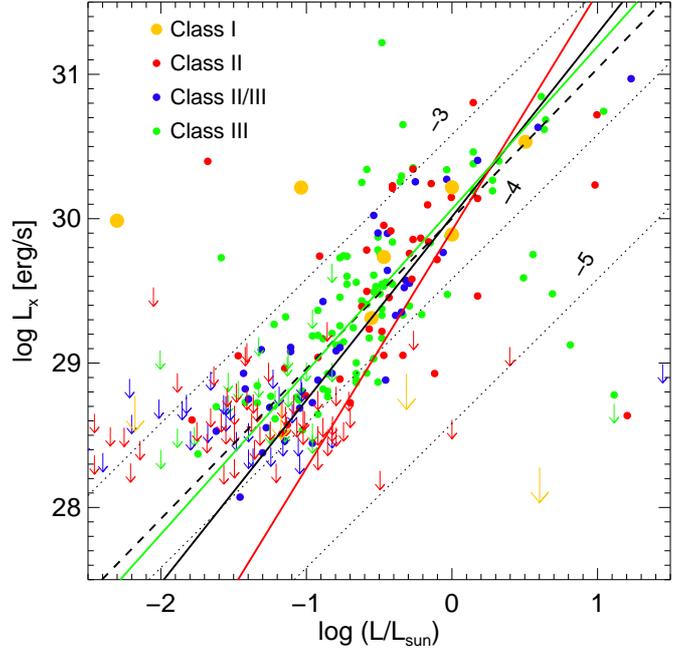}
\caption{
X-ray versus bolometric luminosity (a small number of objects, not relevant for this study, 
falls outside the plotted
$L_{\rm bol}$ range); dotted lines represent constant $L_{\rm x}/L_{\rm bol}$
ratios $10^{-3}$, $10^{-4}$ and $10^{-5}$; the dashed line is the linear regression fit derived 
by \protect\cite{Preibisch05.1} for the ONC 
corresponding to $\log{(L_{\rm x}/L_{\rm bol})} = -3.7$; the solid lines are 
linear regression fits to the IC\,348 members 
in the luminosity range $\log{(L/L_\odot)} = -1.5...+0.5$, 
which corresponds roughly to the upper mass boundary of the 
\protect\cite{Baraffe98.1} models: black, red and green are for the whole sample, only Class\,II
and only Class\,III sources, respectively.}  
\label{fig:lx_lbol}
\end{center}
\end{figure}

The ratio between X-ray and bolometric luminosity 
is thought
to be a crucial indicator for the level of magnetic activity in young stars. 
From X-ray surveys in various star forming regions it has emerged that the 
characteristic emission level of most low-mass pre-main sequence stars is concentrated
between $\log{(L_{\rm x}/L_{\rm bol})} \approx -3$ and $-4$ but with a spread of at least 
$3$ dex, and there are indications for a slight decline of the 
$\log{(L_{\rm x}/L_{\rm bol})}$ level with decreasing mass
\citep[e.g.][]{Flaccomio03.2, Preibisch05.1, Preibisch05.2, Grosso07.1}.

\subsubsection{Sample without distinction of YSO class}\label{subsubsect:results_lbol_all}

Fig.~\ref{fig:lx_lbol} shows the $L_{\rm x}$ vs. $L_{\rm bol}$ diagram 
for IC\,348. 
The sample of known IC\,348 members extends to lower
bolometric luminosities and masses than that of Orion, possibly due to both environmental and observational effects 
(e.g. higher extinction in Orion) and the closer distance.
We carried out a linear regression fit for the range $\log{(L/L_\odot)} = -1.5...+0.5$
using the EM algorithm in ASURV \citep{Feigelson85.1}. 
Our choice for the luminosity cutoffs are motivated by the properties of the IC\,348
sample (avoiding at the high end X-ray faint massive stars outside the mass range covered by
the \cite{Baraffe98.1} models and avoiding to be dominated by upper limits in X-rays at the low
end). 
The result is overplotted in Fig.~\ref{fig:lx_lbol} together with the bestfit obtained
by \cite{Preibisch05.1} for the ONC in a similar range of bolometric luminosities.
The ONC study was performed for stars with 
$\log{(L/L_\odot)} < 1$ and without lower cutoff to the bolometric luminosity. However, 
effectively the ONC sample extends down to $\log{(L/L_\odot)} = -1.5$, i.e. 
the same as our low-luminosity cutoff. 
With respect to \cite{Preibisch05.1}, we have chosen a lower value for the 
luminosity cutoff at the high end because at the age of IC\,348 
stars with $\log{(L/L_\odot)} > 0.5$
are mostly radiative (cf. Fig.~\ref{fig:hrd}) 
and, therefore, their X-ray emission -- if any -- may not have
the same origin as in lower-mass stars. 

The slope we determined for IC\,348 in the luminosity range described above
($1.27 \pm 0.08$) 
is significantly steeper than the one measured from COUP for the ONC ($1.04 \pm 0.06$)
and from XEST for Taurus \cite[$1.05 \pm 0.06$; ][]{Telleschi07.0}.
While in Orion and Taurus the linear fit of $L_{\rm x}$ vs. $L_{\rm bol}$ 
is consistent with a constant $L_{\rm x}/L_{\rm bol}$ ratio, in IC\,348 we observe a 
decrease of $L_{\rm x}/L_{\rm bol}$ with decreasing bolometric luminosity.  
A possible reason may be the particular distribution of YSO types and their different
X-ray luminosities. 
Note, however, that the COUP and XEST samples we refer to 
are not directly equivalent to our samples
in IC\,348 because of different ways the subsamples were defined. 
\cite{Preibisch05.1} distinguish accretors from non-accretors on the basis 
of $8542$\,\AA~ calcium emission. Similarly, \cite{Telleschi07.0} 
separate accretors from non-accretors on the basis of H$\alpha$ emission,
while we have chosen to work with the {\em Spitzer} YSO classification because it yields the
largest sample size. 

\subsubsection{Class\,II vs. Class\,III and comparison with the ONC}\label{subsubsect:results_lbol_yso}

When the same analysis is carried out separately for the subsamples of Class\,II and Class\,III 
sources in IC\,348 we find slopes of $1.65 \pm 0.22$ and $1.13 \pm 0.11$, respectively. 
These results are graphically demonstrated in Fig.~\ref{fig:lx_lbol}. To summarize, we find that 
\begin{itemize}
\item the $L_{\rm x} - L_{\rm bol}$ relation of Class\,III objects in IC\,348 is 
in agreement with that of the non-accretors in the ONC and is roughly constant across the
examined range of bolometric luminosities; 
\item the Class\,II sources in IC\,348 show a steeper slope, i.e. decreasing $L_{\rm x} - L_{\rm bol}$ 
ratio towards fainter stars. 
IC\,348 hosts a large number of Class\,II sources with low luminosities (i.e. low masses) -- 
see e.g. Table~\ref{tab:yso_statistics} -- and these objects have low X-ray luminosities or 
they are not detected. 
\end{itemize}

\begin{figure*}
\begin{center}
\parbox{18cm}{
\parbox{9cm}{
\resizebox{9.cm}{!}{\includegraphics{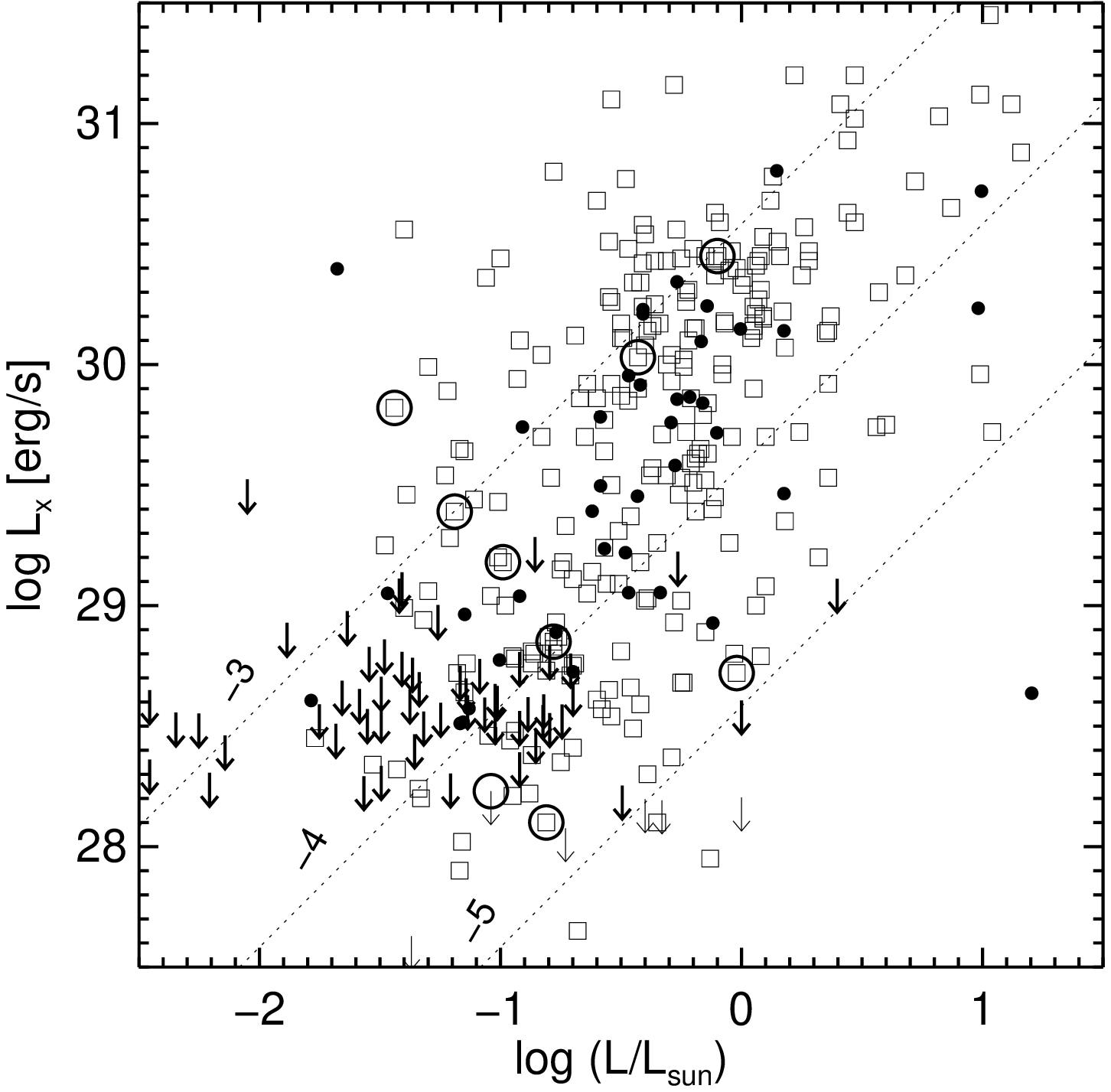}}
}
\parbox{9cm}{
\resizebox{9.cm}{!}{\includegraphics{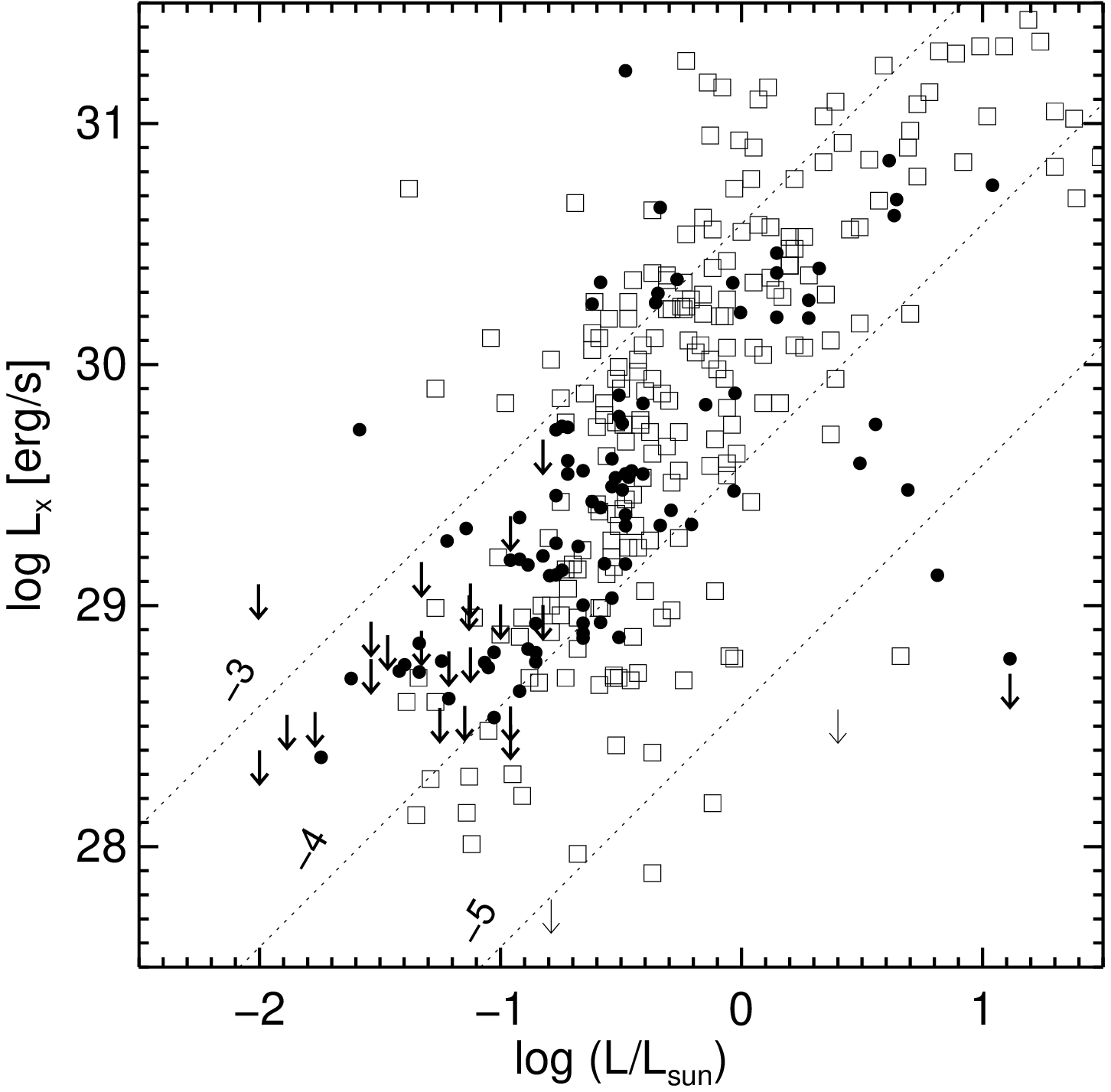}}
}
}
\caption{X-ray versus bolometric luminosity for IC\,348 (filled, thick plotting symbols) 
compared to the ONC (open, thin plotting symbols): 
(left) - Class\,II, (right) - Class\,III. Stars with indications for a disk warp provoking
temporarily high extinction (AA\,Tau-like stars) are highlighted with large circles.}
\label{fig:lx_lbol_pluscoup}
\end{center}
\end{figure*}

Only recently YSO classifications for ONC stars based on {\em Spitzer}/IRAC photometry 
have become available. We extracted these data from the 
{\em YSOVAR database}\footnote{http://ysovar.ipac.caltech.edu}.
In Fig.~\ref{fig:lx_lbol_pluscoup} we compare the X-ray vs. bolometric luminosities for
IC\,348 with those of the ONC separately for Class\,II and Class\,III sources.

\begin{figure}
\begin{center}
\resizebox{9.cm}{!}{\includegraphics{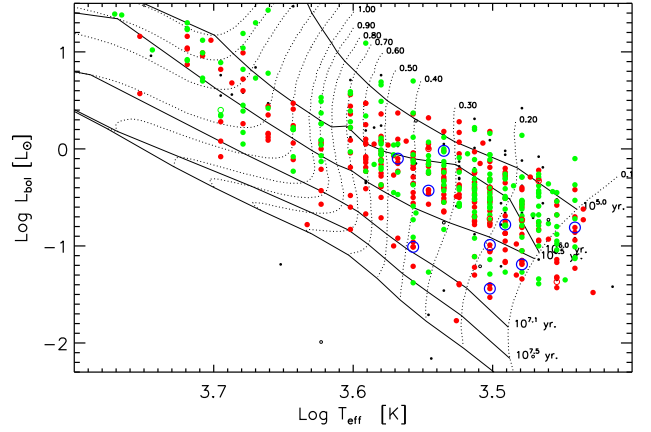}}
\caption{HR diagram for the optical COUP sample on the \cite{Siess00.1} models. 
Class\,III (green; light grey in b/w print) and Class\,II (red; dark grey in b/w print), 
AA\,Tau like stars (large blue annuli) and IC\,348 members without
YSO classification (small black points).}
\label{fig:coup_hrd}
\end{center}
\end{figure}

According to 
Fig.~\ref{fig:lx_lbol_pluscoup}(left), in contrast to IC\,348 
there is no evidence for an X-ray deficiency of low-luminosity diskbearing stars in the ONC. 
However, a look into the HR diagram for the ONC, presented in Fig.~\ref{fig:coup_hrd}, 
suggests that the bolometric luminosities of these stars (i.e. the ones 
with $\log{L_{\rm bol}} \sim -1...-1.5$) 
may be underestimated. This group is characterized by an obvious excess of Class\,II sources 
with isochronal age around $10$\,Myr or more in the HR diagram, in contradiction with 
the general paradigm that
Class\,II stars are on average less evolved than Class\,III stars as well as with
plausible timescales for disk dissipation of just a few Myr. We hypothesize that many
of these low-luminosity, low-mass Class\,II objects are 
either observed in scattered light, the central star being fully obscured by an edge-on disk,
or partially obscured by disk warps or other circumstellar structures
the so-called AA\,Tau phenomenon. A number of AA\,Tau-like stars in the ONC were 
identified by \cite{MoralesCalderon11.0} and are 
marked in Fig.~\ref{fig:coup_hrd} by large annuli. 
Not all of the cluster members with untypically low bolometric luminosity 
for their effective temperature display the AA\,Tau features but this is not in contradiction with
our hypothesis because occultations by disk warps are transient phenomena. In fact, these objects
were identified by \cite{MoralesCalderon11.0} on the basis of their time-variability. 
Moreover, some of these objects may be permanently obscured systems with edge-on disk. 
Both of the described scenarios 
would imply that the extinction correction for these stars has been significantly underestimated,
thus placing the true position of the stars in the $L_{\rm x} - L_{\rm bol}$ diagram 
further to the right 
and in better agreement with the trend that we see in IC\,348. The fact that no such Class\,II
stars with low $L_{\rm bol}$ but high $L_{\rm x}$ are present in IC\,348 might be explained by
the older age with respect to the ONC, going along with the dispersal of the circumstellar
material that provides the extinction. 

We pass now to the consideration of the Class\,III sources (Fig.~\ref{fig:lx_lbol_pluscoup}\,right). 
In contrast to the analysis of \cite{Preibisch05.1} for non-accretors, 
the ONC Class\,III sample shows evidence for a steeper 
$L_{\rm x} - L_{\rm bol}$ relation than the one corresponding to a constant luminosity ratio.
This is in line with the result for Class\,II sources 
considering the hypothesis made above concerning AA\,Tau like stars. 
We could not detect this deficiency of low-luminosity X-ray faint Class\,III sources in IC\,348
but this may be due to the limited sample size. From Fig.~\ref{fig:lx_lbol_pluscoup} it appears
that the IC\,348 and ONC Class\,III distributions are rather similar.


\subsection{X-ray luminosity functions}\label{subsect:results_xlf}

Cumulative distributions of the X-ray luminosities are traditionally used to compare subsamples
of pre-main sequence stars. 
In the past, the X-ray luminosity functions (XLF) derived for various star forming
regions have been vigourously debated, as it could not be established if circumstellar matter 
influences these distributions. Most studies that have distinguished accreting and non-accreting
stars on basis of the strength of their H$\alpha$ emission found that cTTS have
systematically lower X-ray luminosities than wTTS \citep{Stelzer01.1, Flaccomio03.2}. 
On the other hand, no differences were found between stars with and without disks as
diagnosed by IR excess (e.g. PZ01). 

We have constructed cumulative distributions of both the X-ray luminosity (i.e. the XLFs)
and the $L_{\rm x}/L_{\rm bol}$ ratio. 
We computed XLFs for YSOs of different SED classes (to examine the influence of disks) and for
different strength of H$\alpha$ emission (to examine the influence of accretion). 
To avoid biases related to the known dependence of $L_{\rm x}$ on stellar mass, 
the analysis was carried out in three mass bins chosen such that they sample adequately
the low-mass IC\,348 population: $0.1-0.25\,M_\odot$, $0.25-0.6\,M_\odot$, $0.6-1.2\,M_\odot$. 
The significance of the results is verified with two-sample tests that yield the probability 
for rejecting the hypothesis that the
observed luminosity distributions are drawn from the same parent distributions. 
Both the cumulative distributions and the two-sample tests are computed within the 
ASURV environment. 
The calculations take into account the upper limits for undetected stars. 
 
\begin{figure*}
\begin{center}
\parbox{18cm}{
\parbox{9cm}{
\resizebox{9.5cm}{!}{\includegraphics{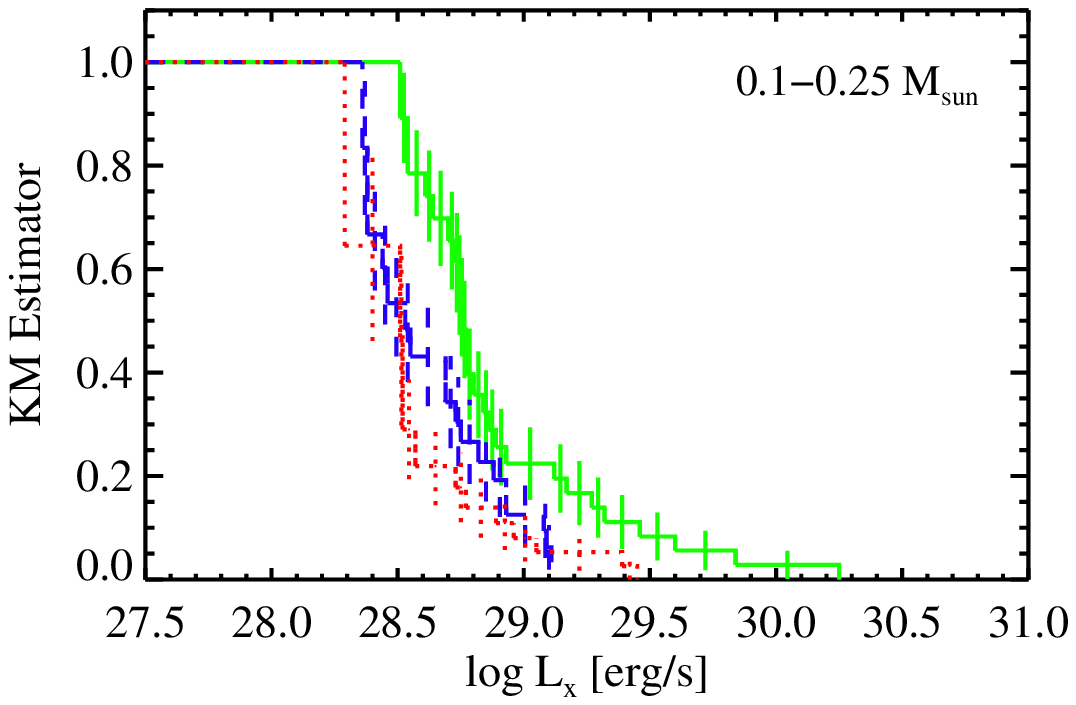}}
}
\parbox{9cm}{
\resizebox{9.5cm}{!}{\includegraphics{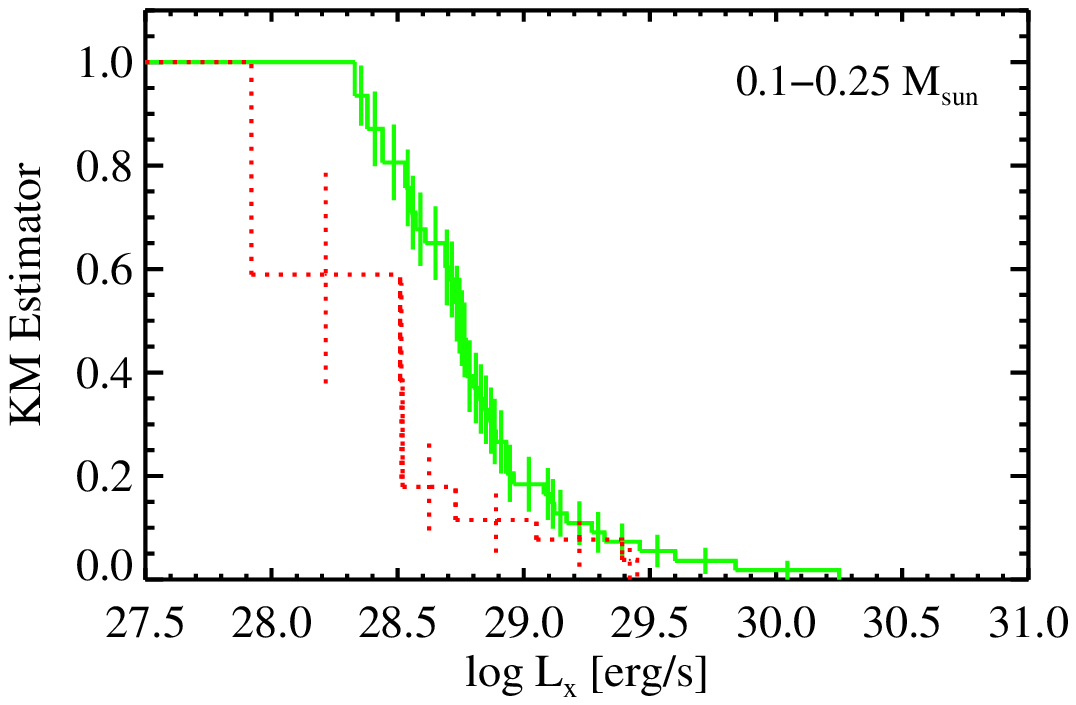}}
}
}
\parbox{18cm}{
\parbox{9cm}{
\resizebox{9.5cm}{!}{\includegraphics{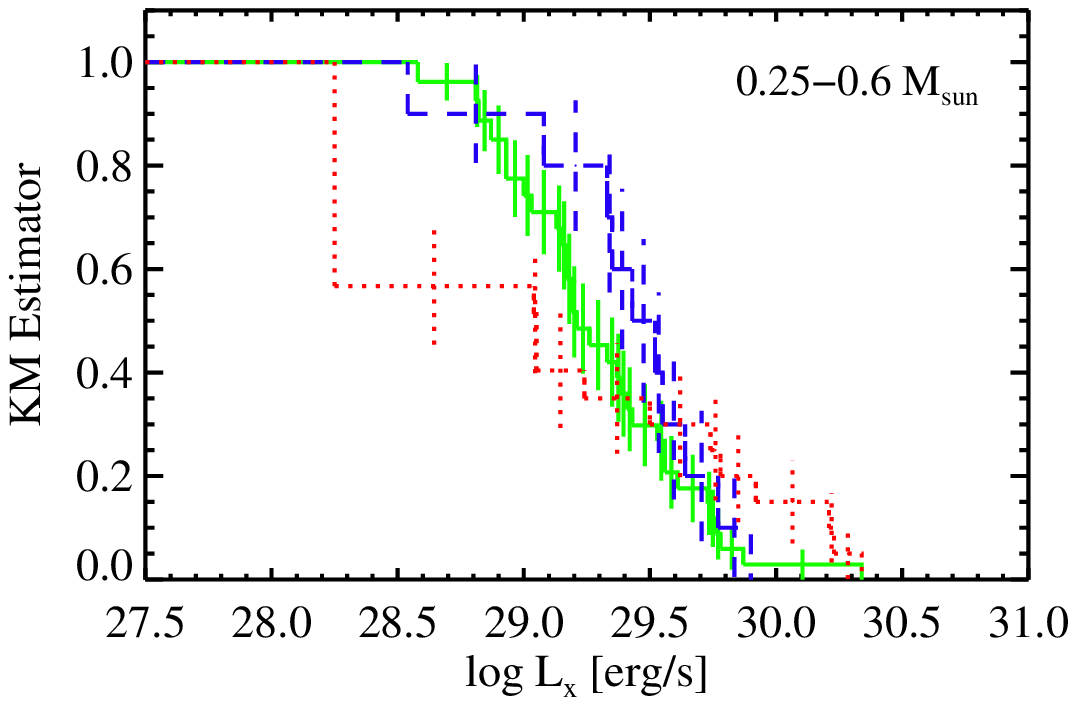}}
}
\parbox{9cm}{
\resizebox{9.5cm}{!}{\includegraphics{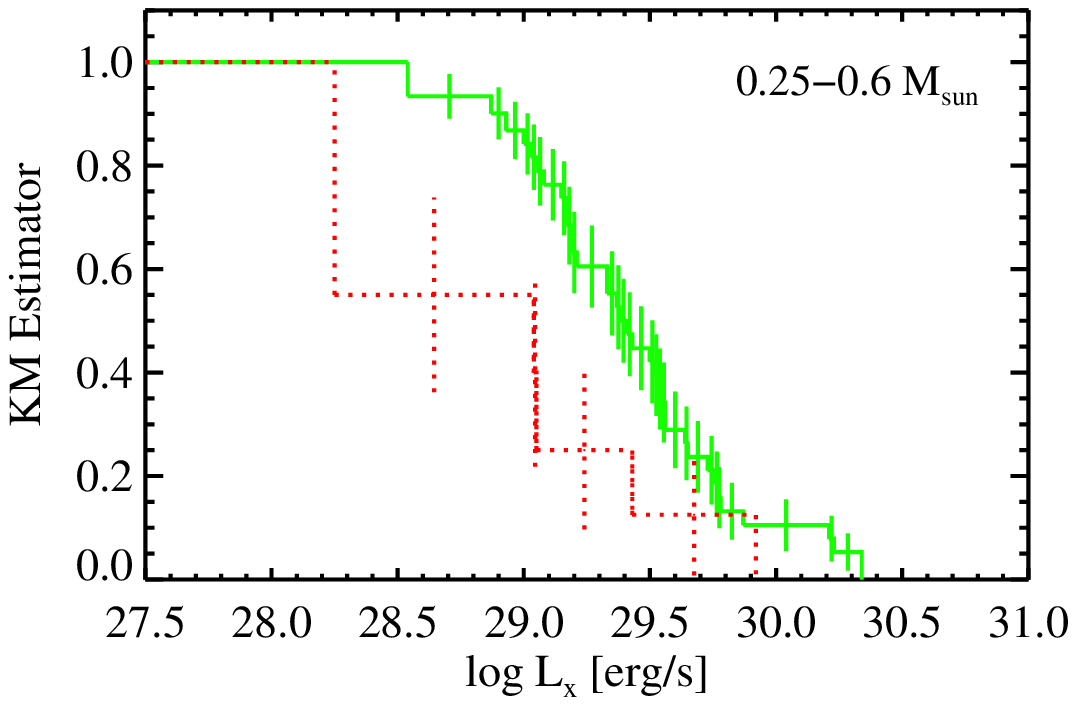}}
}
}
\parbox{18cm}{
\parbox{9cm}{
\resizebox{9.5cm}{!}{\includegraphics{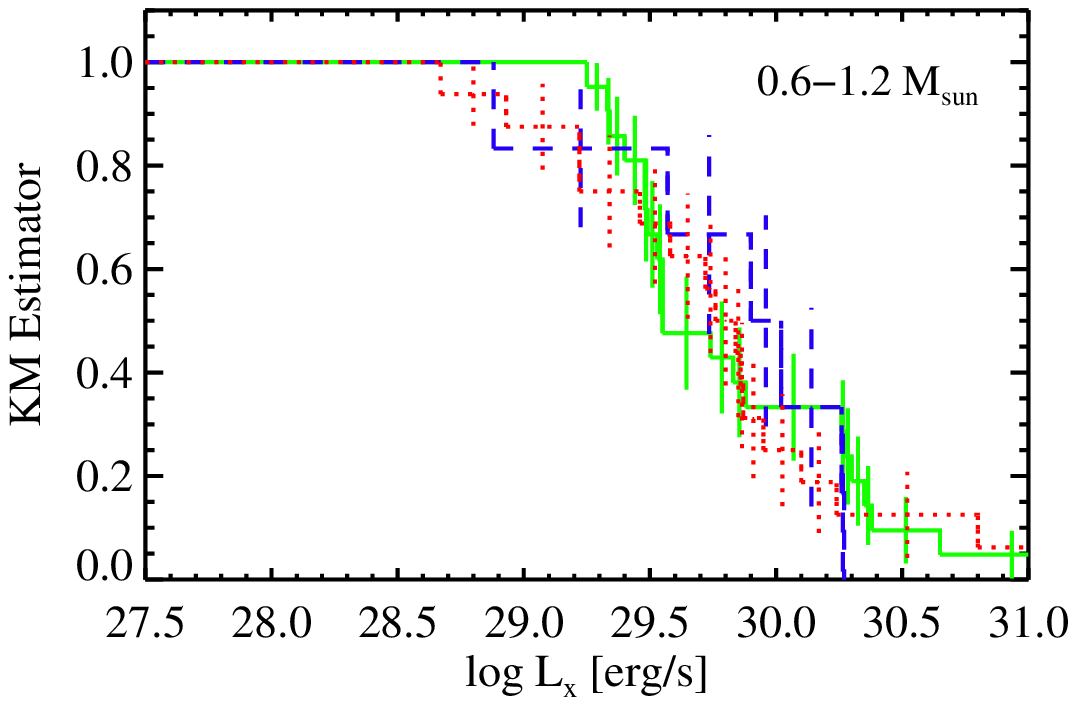}}
}
\parbox{9cm}{
\resizebox{9.5cm}{!}{\includegraphics{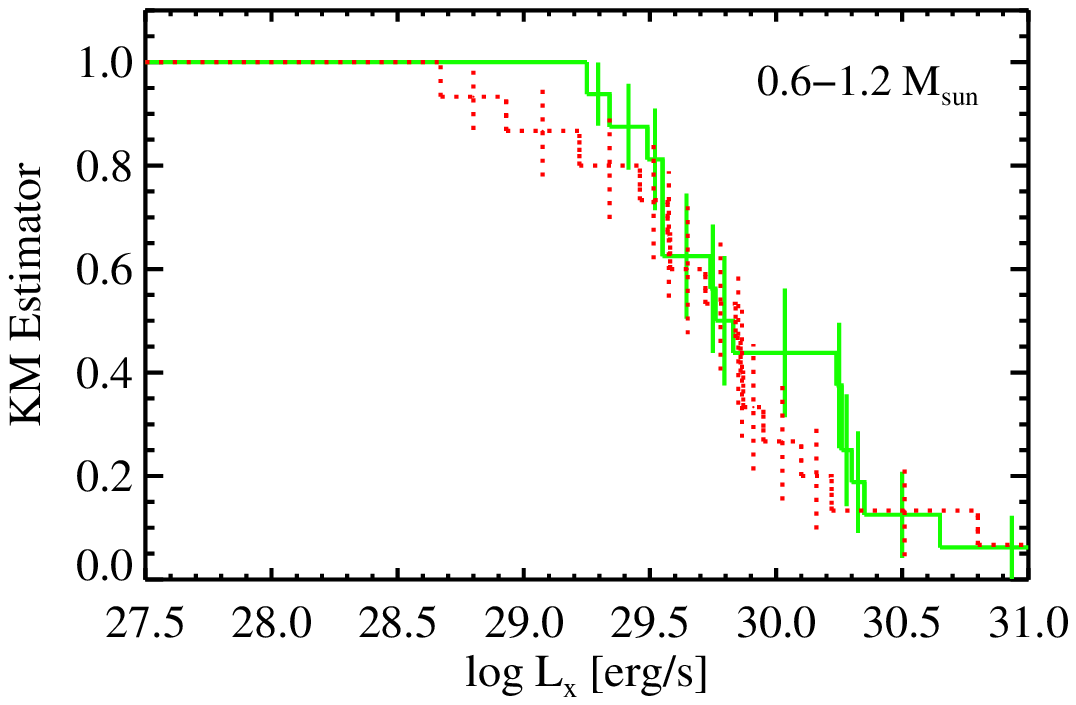}}
}
}
\caption{Cumulative distributions of X-ray luminosity 
for different subgroups of YSO in three mass bins (from top to bottom). 
The left panels represent three YSO classes based on IR excess 
(red dotted - Class\,II, blue dashed - Class\,II/III, and green solid - Class\,III)
and the right panels represent YSO classes based on H$\alpha$ emission 
(green solid - wTTS and red dotted - cTTS).}
\label{fig:xlf}
\end{center}
\end{figure*}

\begin{figure*}
\begin{center}
\parbox{18cm}{
\parbox{9cm}{
\resizebox{9.5cm}{!}{\includegraphics{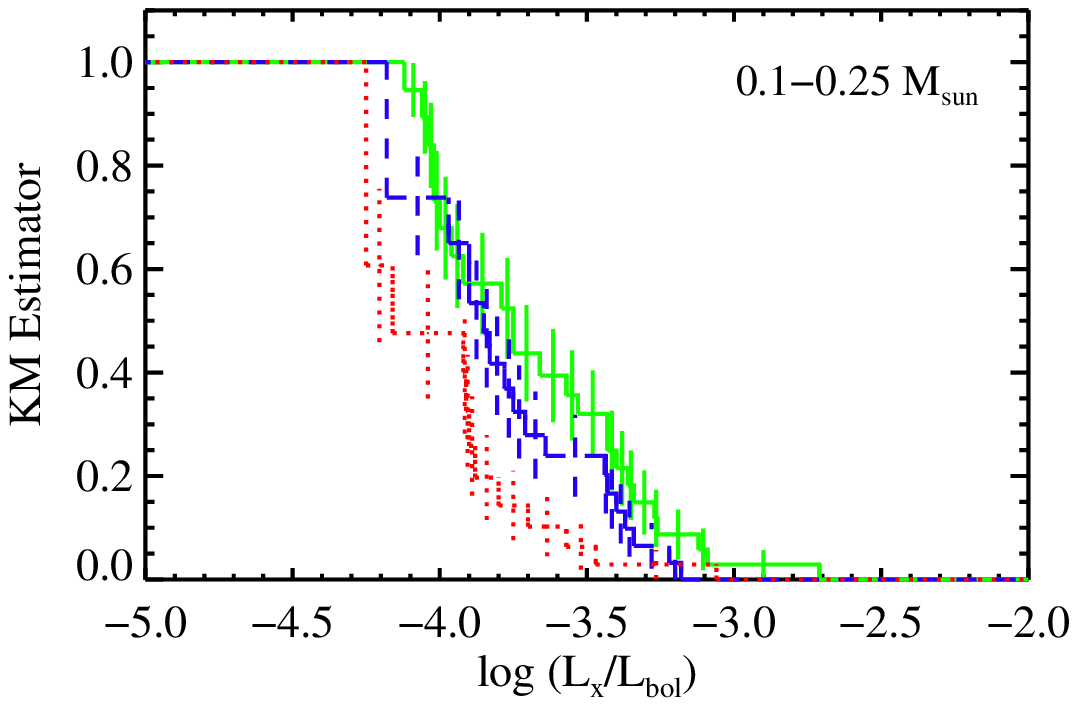}}
}
\parbox{9cm}{
\resizebox{9.5cm}{!}{\includegraphics{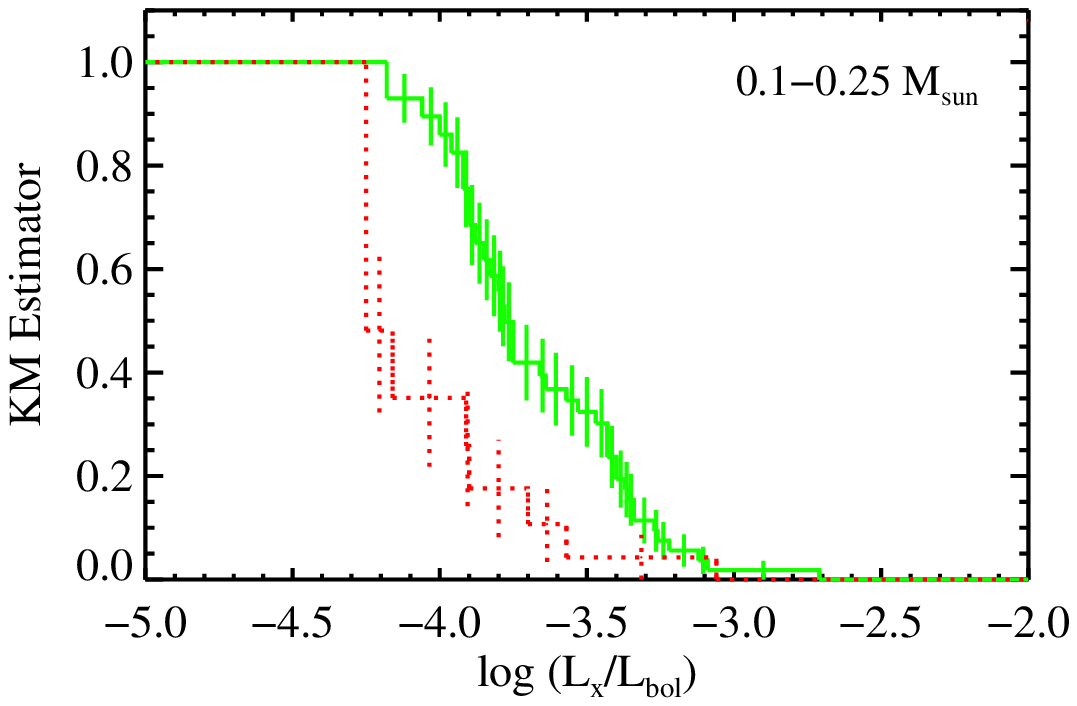}}
}
}
\parbox{18cm}{
\parbox{9cm}{
\resizebox{9.5cm}{!}{\includegraphics{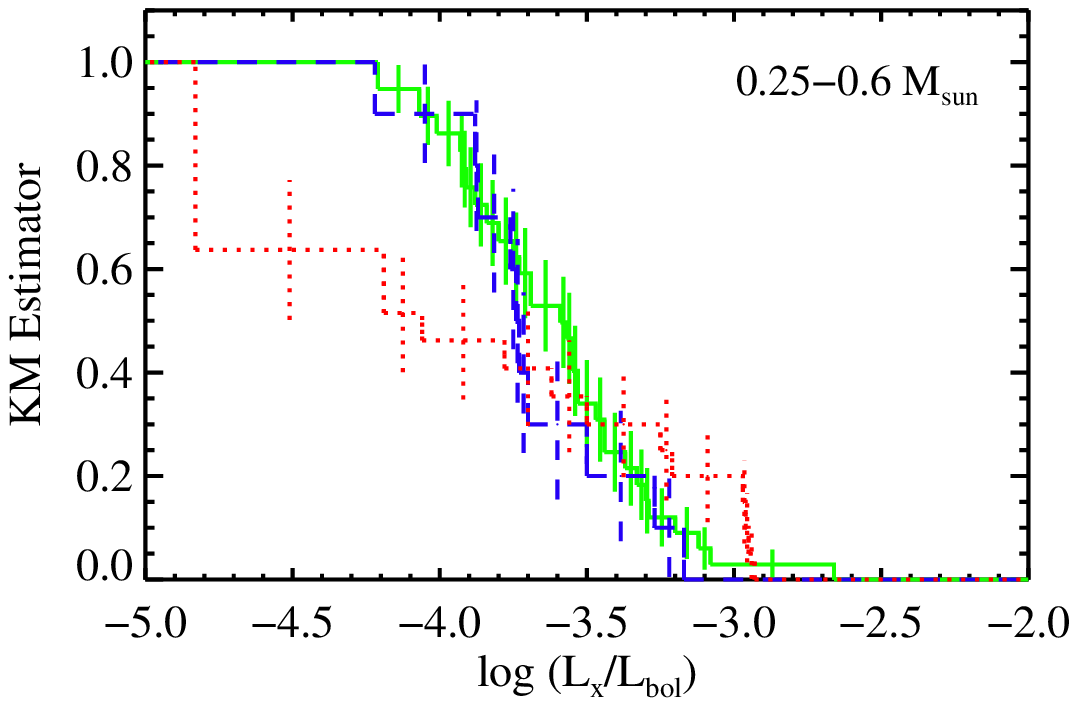}}
}
\parbox{9cm}{
\resizebox{9.5cm}{!}{\includegraphics{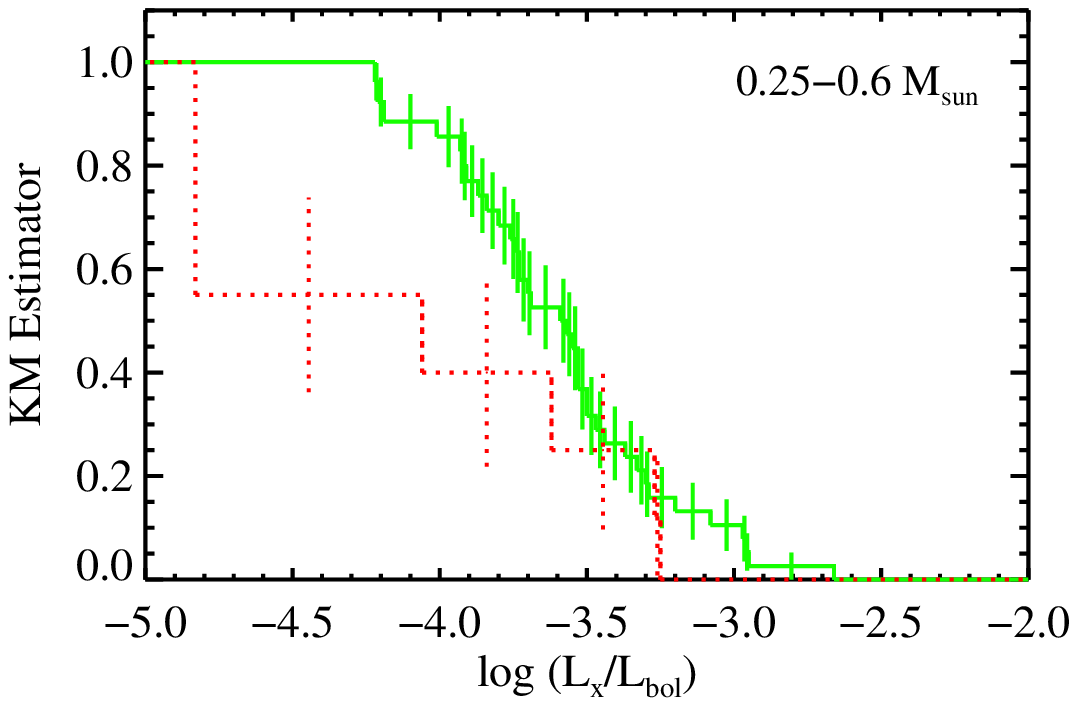}}
}
}
\parbox{18cm}{
\parbox{9cm}{
\resizebox{9.5cm}{!}{\includegraphics{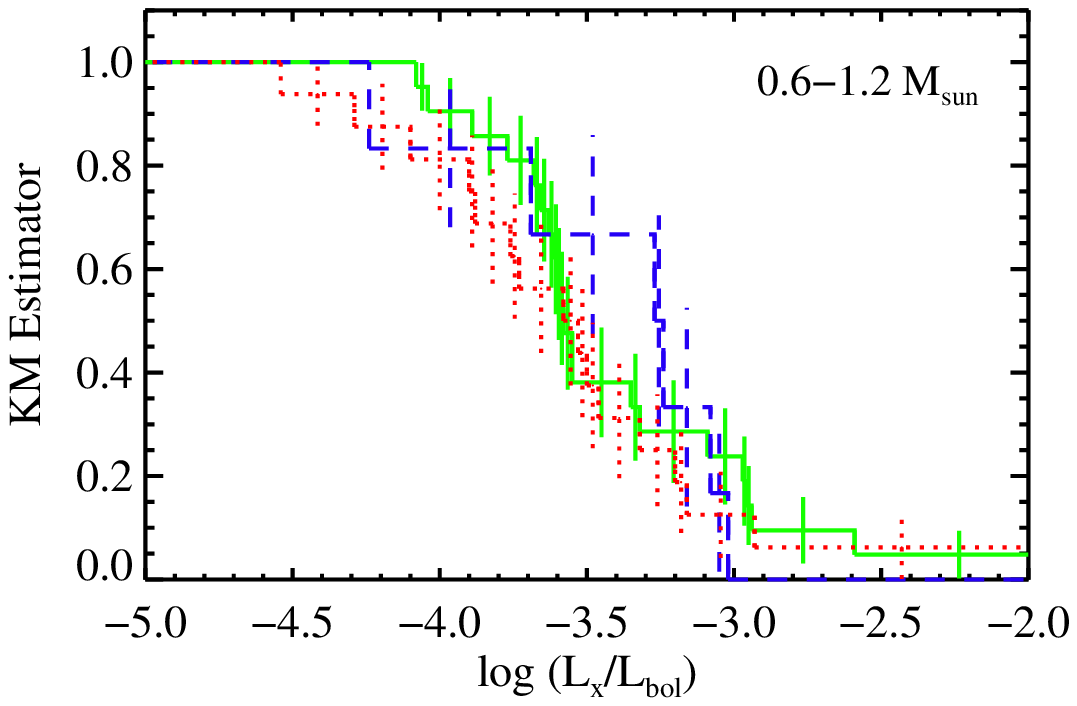}}
}
\parbox{9cm}{
\resizebox{9.5cm}{!}{\includegraphics{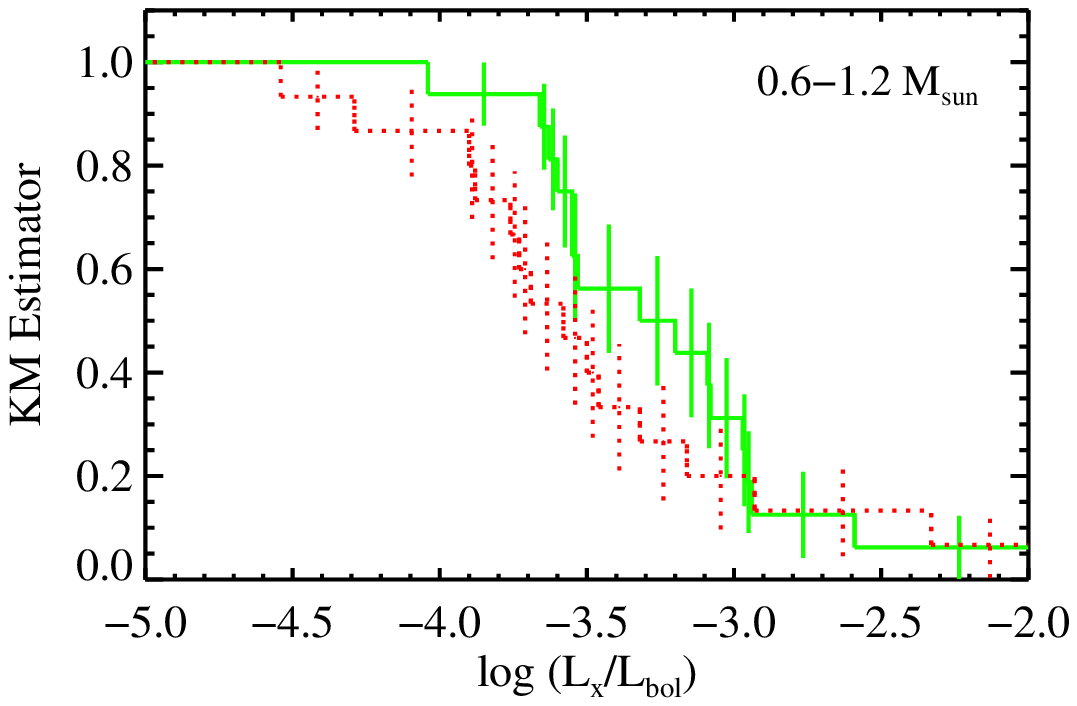}}
}
}
\caption{Same as Fig.~\ref{fig:xlf} but for the fractional X-ray luminosity
$\log{(L_{\rm x}/L_{\rm bol})}$.} 
\label{fig:xlf_lxlbol}
\end{center}
\end{figure*}

The results are displayed in Figs.~\ref{fig:xlf} and~\ref{fig:xlf_lxlbol}. In the left
hand panels 
we compare the XLFs of Class\,II, II/III, and III sources.
Class\,I objects are not considered because of poor statistics. 
There is a pronounced difference in the X-ray luminosities of the three YSO classes for
the lowest mass bin ($0.1-0.25\,M_\odot$), where Class\,III sources have higher $L_{\rm x}$ than
both Class\,II/III and Class\,II and the latter two classes are indistinguishable. 
This is also the mass range with the highest number statistics comprising $106$ stars.
However, these low-mass samples have a large number of upper limits. 
A marginal difference between Class\,II objects and the other two YSO groups 
is found for the intermediate mass bin
($0.25-0.6\,M_\odot$) only in the logrank test 
while all other
types of two-sample tests could not detect a difference with better than $90$\,\% probability. 
For the highest mass bin ($0.6-1.2\,M_\odot$) the XLFs of the three types of YSOs can not
be distinguished. 
The graphical results are supported by the two-sample tests. 
The statistics are summarized in Table~\ref{tab:xlf_yso}
where  $'N$' is the total number of stars and $'N_{\rm ul}'$ is the number of 
undetected stars. The luminosities refer to the median of each sample, 
and the numbers in the last three columns represent probabilities 
for two distributions being different obtained from two-sample tests.
\begin{table*}
\begin{center}
\caption{Statistics for detections, median of the logarithm of X-ray luminosity 
and $L_{\rm x}/L_{\rm bol}$ ratio, and $2$-sample tests of different YSO classes}
\label{tab:xlf_yso}
\begin{tabular}{lrrrrrrrrr}
\hline
                     &                  &                           &                &                             &                 &                            & 2S-Test         & 2S-Test             & 2S-Test            \\
                     & \multicolumn{2}{c}{Class\,III}               & \multicolumn{2}{c}{Class\,II/III}            & \multicolumn{2}{c}{Class\,II}                & Class\,III - II & Class\,III - II/III & Class\,II/III - II \\
$M\,{\rm [M_\odot]}$ & $N/N_{\rm ul}$ & $\log{L_{\rm x}}$ & $N/N_{\rm ul}$ & $\log{L_{\rm x}}$ & $N/N_{\rm ul}$ & $\log{L_{\rm x}}$ & Prob.[\%]       & Prob.[\%]           & Prob.[\%]          \\
\hline
$0.1-0.25$  & $36/13$ & $28.76$ & $32/15$ & $28.51$ & $38/28$ & $28.47$ & $>99$ & $>97$ & $<90$ \\
$0.25-0.6$  & $34/4$  & $29.20$ & $10/1$  & $29.43$ & $20/9$  & $29.04$ & $<90$\tablefootmark{a} & $<90$ & $<90$\tablefootmark{a} \\
$0.6-1.2$   & $21/0$  & $29.55$ & $6/0$   & $29.90$ & $16/0$  & $29.76$ & $<90$ & $<90$ & $<90$ \\
\hline
$M\,{\rm [M_\odot]}$ & $N/N_{\rm ul}$ & $\log{L_{\rm x}/L_{\rm bol}}$ & $N/N_{\rm ul}$ & $\log{L_{\rm x}/L_{\rm bol}}$ & $N/N_{\rm ul}$ & $\log{L_{\rm x}/L_{\rm bol}}$ & Prob.[\%]       & Prob.[\%]           & Prob.[\%]          \\
\hline
$0.1-0.25$  & $36/13$ & $-3.78$ & $32/15$ & $-3.87$ & $38/28$ & $-4.18$ & $>99$ & $<90$ & $>97$ \\
$0.25-0.6$  & $34/4$  & $-3.60$ & $10/1$  & $-3.74$ & $20/9$  & $-4.15$ & $<90$\tablefootmark{b} & $<90$ & $<90$ \\
$0.6-1.2$   & $21/0$  & $-3.59$ & $6/0$   & $-3.27$ & $16/0$  & $-3.58$ & $<90$ & $<90$ & $<90$ \\
\hline
\end{tabular}
\tablefoot{
\tablefoottext{a}{The logrank test gives probabilities of $>97$\,\% and $>93$\,\%
for Class\,III vs. Class\,II and Class\,II/III vs. Class\,II.}
\tablefoottext{b}{The logrank test gives a probability of $>99$\,\% 
for Class\,II vs. Class\,II.}
}
\end{center}
\end{table*}

Also in terms of $L_{\rm x}/L_{\rm bol}$ there is a marked difference between the three
YSO classes only in the lowest mass bin.
In contrast to the XLFs, where Class\,II/III objects behave similar to the Class\,II sources, 
in $L_{\rm x}/L_{\rm bol}$ the Class\,II/III objects are more similar to the Class\,III sources
(compare uppermost left panels in Figs.~\ref{fig:xlf} and~\ref{fig:xlf_lxlbol}).  
Again only the logrank test gives Class\,II objects different from Class\,III for the
intermediate mass bin. 
All other distributions of 
$L_{\rm x}/L_{\rm bol}$ for higher masses are undistinguishable.
These results confirm our findings from Fig.~\ref{fig:lx_lbol} that the $L_{\rm x} - L_{\rm bol}$
relation in IC\,348 is steeper for Class\,II than for Class\,III sources.  

An analogous investigation was performed for cTTS and wTTS, discriminated by the 
equivalent width of H$\alpha$ as described in Sect.~\ref{subsect:cat_derived}.  
The XLFs and the cumulative distributions for $L_{\rm x}/L_{\rm bol}$  
for cTTS and wTTS are shown in the right panels of 
Figs.~\ref{fig:xlf} and~\ref{fig:xlf_lxlbol} for the same mass bins 
that we have examined for the samples classified by their IR excess. 
Again, a significant difference between the groups is found for the lowest 
mass bin with wTTS being more X-ray luminous than cTTS; see Table~\ref{tab:xlf_wha}
for a summary of the results. 
In the intermediate mass bin, where the Class\,II stars were different from the other
two samples only in the logrank test, the cTTS and the wTTS are different at high significance
for all two-sample tests if $L_{\rm x}$ is considered but, again, only in the logrank test
if $L_{\rm x}/L_{\rm bol}$ is considered. 

Note, that the IR excess and the H$\alpha$ emission samples are not identical 
and their comparison is not straightforward. 
The number of stars with measured $\alpha_{\rm 3-8\,\mu m}$ is larger than those with H$\alpha$ data. 
For the lowest mass bin, most of the Class\,II/III sources belong to the 
wTTS group. However, there are also several Class\,II sources that are 
classified as non-accretors.
We have shown that the trend of a mass dependence of the X-ray emission level throughout 
the various evolutionary stages of YSOs is robust against details of the sample definition. 
\begin{table}
\begin{center}
\caption{Statistics for detections, median of the logarithm of X-ray luminosity 
and $L_{\rm x}/L_{\rm bol}$ ratio, and two-sample tests for cTTS and wTTS}
\label{tab:xlf_wha}
\begin{tabular}{lrrrrr}
\hline
                     &                &                             &                 &                            & 2S-Test         \\
                     & \multicolumn{2}{c}{wTTS}                     & \multicolumn{2}{c}{cTTS}                     & wTTS - cTTS     \\
$M\,{\rm [M_\odot]}$ & $N/N_{\rm ul}$ & $\log{L_{\rm x}}$ & $N/N_{\rm ul}$ & $\log{L_{\rm x}}$ & Prob.[\%]       \\
\hline
$0.1-0.25$  & $55/18$ & $28.74$ & $26/20$ & $28.18$ & $>99$ \\
$0.25-0.6$  & $38/3$  & $29.38$ & $8/4$   & $28.51$ & $>95$ \\
$0.6-1.2$   & $16/0$  & $29.76$ & $15/0$  & $29.78$ & $<90$ \\
\hline
$M\,{\rm [M_\odot]}$ & $N/N_{\rm ul}$ & $\log{\frac{L_{\rm x}}{L_{\rm bol}}}$ & $N/N_{\rm ul}$ & $\log{\frac{L_{\rm x}}{L_{\rm bol}}}$ & Prob.[\%]       \\
\hline
$0.1-0.25$  & $55/18$ & $-3.78$ & $26/20$ & $-4.09$ & $>99$ \\
$0.25-0.6$  & $38/3$  & $-3.59$ & $8/4$   & $-4.57$ & $<90$\tablefootmark{a} \\
$0.6-1.2$   & $16/0$  & $-3.32$ & $15/0$  & $-3.64$ & $<90$ \\
\hline
\end{tabular}
\tablefoot{
\tablefoottext{*}{The logrank test gives a probability of $>96$\,\%.}
}
\end{center}
\end{table}

\subsection{X-ray luminosity and mass accretion}\label{subsect:results_mdot}

In the past, differences between the X-ray luminosities of cTTS and wTTS have been
ascribed to a possible connection between X-ray emission and accretion.
The interpretations have involved opposite views: 
In one scenario X-ray emission is suppressed by accretion, 
e.g. by a direct influence of accretion onto the dynamo efficiency through  
reduced convection or through reduced coronal heating as a result of the 
higher plasma densities involved in the accretion 
columns \citep[e.g.][]{Preibisch05.1} or by occulting the X-ray emitting
corona \citep{Gregory07.1}. 
In the alternative picture X-ray emission determines whether accretion takes place
or not through photoevaporation of disks. This latter hypothesis has been bolstered by
an empirical anticorrelation of X-ray luminosity and mass accretion rate observed
in COUP \citep{Drake09.1}. 

\begin{figure}
\begin{center}
\resizebox{9.5cm}{!}{\includegraphics{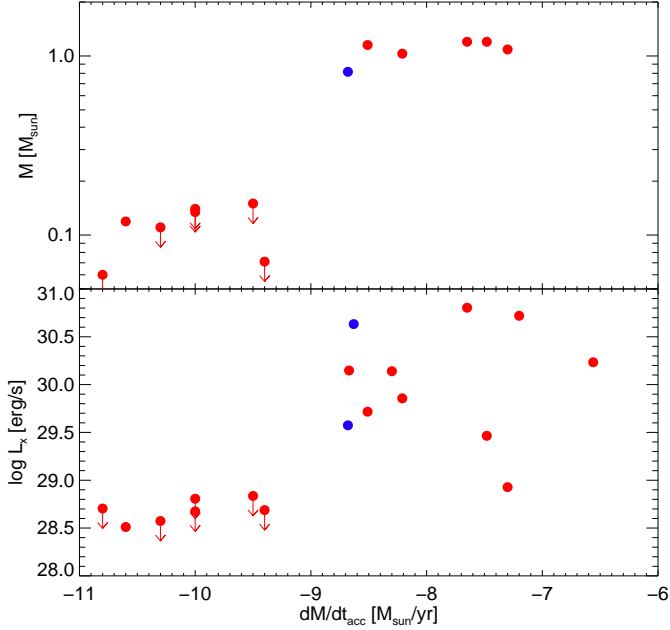}}
\caption{X-ray luminosity and mass versus mass accretion rate. Colors define YSO classes as in 
Fig.~\ref{fig:wha_teff}.}
\label{fig:lx_mdot}
\end{center}
\end{figure}

Only $18$ IC\,348 members have measured mass accretion rate. 
In Fig.~\ref{fig:lx_mdot}  
we present the relation between $L_{\rm x}$ and $\dot{M}_{\rm acc}$ for IC\,348. 
Note, that at least two stars in IC\,348 with `anemic' disks are accreting. 
The top panel of this figure shows that 
the apparent positive correlation between $L_{\rm x}$ and $\dot{M}_{\rm acc}$ 
is a result of the well-known dependence of mass accretion rate on stellar mass 
\cite[see e.g.][]{Natta06.1}. Three of the objects in the group of strong
accretors (empirical dividing line of $\log{\dot{M}_{\rm acc}} = -9$) have no
mass assigned because they are outside the mass range covered by the \cite{Baraffe98.1}
models ($M > 1.2\,M_\odot$). All in all, however, all stars in this group have
similar mass but span a range of two decades in accretion rates and more than one
decade in X-ray luminosity. 
The Spearman's test does not yield evidence for a correlation between the
two parameters for the IC\,348 members with strong accretion. However, the small sample
size precludes any firm conclusions. 

\section{Discussion and conclusions}\label{sect:discussion}

Our combined analysis of all existing {\em Chandra} observations provides
the so far deepest X-ray survey for IC\,348. 
It is difficult to give a number for the sensitivity limit
due to the partial overlap of the available images. 
The exposure time for the deepest part of the merged mosaic is $180$\,ksec and 
the faintest X-ray detected cluster member has a luminosity of 
$\log{L_{\rm x}}\,{\rm [erg/s]} = 28.1$ 
Our detection rate for BDs in IC\,348 is $10$\,\%, in an area surveyed by {\em Chandra} 
that misses only $2$ out of $41$ substellar objects. 

The magnetic activity at and beyond 
the low-mass end of the  main-sequence is generally poorly constrained 
due to a lack of sensitive enough observations in the X-ray regime. 
Even COUP, 
the so far deepest X-ray image of a star formation region, 
has yielded detections for only $26$\,\% of the known BDs, presumably 
due to the considerable extinction in the ONC. 
We caution that the COUP BD sample may be contaminated by a few objects
with mass above the substellar limit due to the fact that the BD classification 
was based mostly on near-IR spectra that are known to yield systematically
later spectral types compared to optical spectra \citep[see][for more details on the 
sample definition]{Preibisch05.2}. 
Nine BDs were detected in Taurus during the XEST 
\citep{Grosso07.1} using the same definition for
BDs as we do in this paper for IC\,348 (spectral type M6 and later).
The XEST sample comprised $17$ BDs known at the time, i.e. the X-ray detection rate was
about $50$\,\%. 
It is important to note that 
many of the X-ray detected BDs in Taurus have untypically high bolometric luminosities
placing them above the $1$\,Myr isochrone of the \cite{Baraffe98.1} models 
\cite[see e.g. Fig.2 in][]{Grosso07.1}. As a consequence, for a given $L_{\rm x}/L_{\rm bol}$ level 
their X-ray luminosity is higher than for IC\,348 BDs explaining the higher detection rate. 

We have examined cumulative distributions of absolute and fractional X-ray luminosity 
in three mass ranges: $0.1-0.25$, $0.25-0.6$, and $0.6-1.2\,M_\odot$. 
The choice of these mass bins was driven by the requirement to have sufficient
stars in each subsample for a statistical analysis. 
The well-known increase of the median X-ray luminosity with increasing mass underlines 
once more that the discrimination of different mass bins is mandatory. 
However, the details of the derived XLFs also depend on the choice of the boundaries for
the mass bins, and more generally on the pre-main sequence models. 
For this reason, results for different star forming regions are difficult to compare. 
In COUP, e.g., the \cite{Siess00.1} models were used,
and the mass bins that were chosen are different from ours motivated by the high number of stars with 
solar mass and above. 

In addition to the issue with stellar masses, the literature provides differing 
definitions for characterizing the evolutionary stage of pre-main sequence stars. 
In COUP, e.g., disks were characterized by near-IR excess
effectively biased towards warmer disks with respect to our {\em Spitzer} YSO classification.
Other COUP subsamples represent accreting and non-accreting stars separated by the
strength of calcium emission. 
In XEST YSOs were distinguished into accretors and non-accretors on the basis of H$\alpha$ 
emission while no disk-status was assigned to the stars.
We have examined for IC\,348 both the influence of the presence of disks and of ongoing accretion
on the X-ray emission level. Our results hold for both types of classifications. 
In particular, we find that differences exist between groups of YSOs in different evolutionary stages
at low masses while they disappear at higher mass. 
This confirms a trend that was seen in COUP \citep{Preibisch05.1}.  
An opposite tendency was claimed in XEST: 
different XLFs for cTTS and wTTS at $M> 0.3\,M_\odot$ but not for lower masses 
\citep{Telleschi07.0}. This result does not hold for IC\,348 according to our data. 

The trends that we see in the XLFs of IC\,348 directly reflect the deficiency of X-ray luminosity
in very low-mass disk-bearing stars 
responsible for the drop of the $L_{\rm x}/L_{\rm bol}$ ratio towards fainter, less massive stars. 
We have compared this result to ONC samples constructed by combining 
the recent YSO classification with COUP-derived parameters, and find a decline of the 
$L_{\rm x}/L_{\rm bol}$ level towards low-luminosity for the diskless stars in the ONC. 
In disk-bearing stars of the ONC the scatter is large, possibly due to biases 
related to the difficulty of correcting the observed luminosities for high
circumstellar extinction that is ubiquitous at the young age of Orion. 

Some scenarios for a decrease of $L_{\rm x}/L_{\rm bol}$ level towards lower masses are: 
(i) transition from a solar-like to a convective dynamo, combined with a lower efficiency 
of the latter one resulting in decreased X-ray production;
(ii) smaller co-rotation radius for lower-mass stars and the ensuing 
centrifugal disruption of the corona \citep{Jardine06.1}; 
(iii) a mass dependent fraction of stars with X-ray emission suppressed by accretion 
\citep{Flaccomio03.2}. 
Concerning (ii), we note that indeed the IC\,348 stars in our lowest mass bin have all short
rotation periods ($\leq 5$\,d) while the period distribution is much broader for higher mass
stars, an issue discussed by \cite{Cieza06.1}. 
Alexander et al. (in prep.) present a detailed study of the rotation/activity connection in
IC\,348 and find no significant differences between stars in the saturated and in the super-saturated
regime. 
Moreover, rotational effects should affect both disk-bearing and diskless stars in the same manner. 
Scenario (iii) is the only one that distinguishes between YSOs of given mass but in 
different evolutionary phases and would explain the different XLFs 
of very low-mass Class\,II and Class\,III objects.
The search for a direct connection between mass accretion rate and X-ray luminosity
could shed light on this question but, at present, it 
is hampered because of the small number of IC\,348 members with accretion measures. In the available
sample of $12$ objects we find no relation between $\dot{M}_{\rm acc}$ and $L_{\rm x}$,
except for the trend that can be ascribed to the well-known dependence of both parameters on mass.

To summarize, 
we have examined one of the deepest available X-ray data sets on a young stellar cluster. 
Thanks to its rich population of very low-mass cluster members 
IC\,348 provides an excellent opportunity to study the X-ray activity level at the boundary
and beyond the substellar limit.  
We have established for the first time with statistical significance 
the non-constant behavior of $L_{\rm x}/L_{\rm bol}$ for the very low-mass pre-main sequence. 
Although the sensitivity of the X-ray data is comparable to that of other recent
studies of star forming regions such as XEST and COUP, 
the well-characterized cluster census way down into the BD regime results in a low
X-ray detection rate for very low-mass objects and  
the low-luminosity end of the X-ray data of IC\,348 continues to be dominated 
by insufficiently constrained upper limits. 
A much longer exposure would 
substantially increase the X-ray census at the low-mass end of the stellar
sequence in IC\,348. 
Additional studies of accretion rates in IC\,348 would 
help to understand the interplay between accretion and X-ray emission.

\begin{acknowledgements} 
We would like to thank the referee, M.G\"udel. 
BS, GM, EF and SS acknowledge financial contribution from the agreement ASI-INAF I/009/10/0.
This work was also supported by the Munich Cluster of Excellence `Origin and Structure of
the Universe'.
\end{acknowledgements}

\bibliographystyle{aa} 
\bibliography{ic348}

\clearpage

%
%
%
%

\begin{deluxetable}{rcrrrrrrrrrrrccrr}
\centering \rotate \tabletypesize{\tiny} \tablewidth{0pt}

\tablecaption{Primary {\em Chandra} Catalog:  Basic Source Properties \label{tab:src_properties_main}}

\tablehead{
\multicolumn{2}{c}{Source} &
\multicolumn{4}{c}{Position} &
\multicolumn{5}{c}{Extraction} &
\multicolumn{6}{c}{Characteristics} \\
                                
\multicolumn{2}{c}{\hrulefill} &  
\multicolumn{4}{c}{\hrulefill} &
\multicolumn{5}{c}{\hrulefill} &
\multicolumn{6}{c}{\hrulefill} \\

\colhead{Seq. No.} & \colhead{CXOU J} &
\colhead{$\alpha$ (J2000.0)} & \colhead{$\delta$ (J2000.0)} & \colhead{Error} & \colhead{$\theta$} &
\colhead{$C_{t,net}$} & \colhead{$\sigma_{t,net}$} & \colhead{$B_{t}$} & \colhead{$C_{h,net}$} & \colhead{PSF Frac.} &   
\colhead{Signif.} & \colhead{$\log P_B$} & \colhead{Anom.} & \colhead{Var.} &\colhead{Eff. Exp.} & \colhead{$E_{median}$}  \\

\colhead{} & \colhead{} &
\colhead{(\arcdeg)} & \colhead{(\arcdeg)} & \colhead{(\arcsec)} & \colhead{(\arcmin)} &
\colhead{(counts)} & \colhead{(counts)} & \colhead{(counts)} & \colhead{(counts)} & \colhead{} &
\colhead{} & \colhead{} & \colhead{} & \colhead{} & \colhead{(ks)} & \colhead{(keV)}
 \\

\colhead{(1)} & \colhead{(2)} &
\colhead{(3)} & \colhead{(4)} & \colhead{(5)} & \colhead{(6)} &
\colhead{(7)} & \colhead{(8)} & \colhead{(9)} & \colhead{(10)} & 
\colhead{(11)} &
\colhead{(12)} & \colhead{(13)} & \colhead{(14)} & \colhead{(15)} & 
\colhead{(16)} & \colhead{(17)}}



\startdata
   1 & 034313.75$+$320045.2 & 55.807303 &  32.012563 &     0.7 &    10.4 &   50.0 &     9.5 &    27.0 &    27.5 &    0.91 &    5.0 & $<$-5 & .... & b &   100.4 &     2.1 \\
   2 & 034316.32$+$320305.4 & 55.818019 &  32.051512 &     0.7 &    10.8 &   65.4 &    10.4 &    27.6 &    54.4 &    0.90 &    6.0 & $<$-5 & .... & a &   100.1 &     3.5 \\
   3 & 034319.89$+$320241.4 & 55.832910 &  32.044855 &     0.8 &    10.0 &   27.9 &     7.4 &    17.1 &     4.4 &    0.90 &    3.5 & $<$-5 & g... & \nodata &    93.5 &     1.3 \\
   4 & 034326.96$+$320042.1 & 55.862341 &  32.011705 &     0.5 &     7.8 &   34.3 &     7.1 &     8.7 &     3.5 &    0.90 &    4.5 & $<$-5 & .... & a &   111.4 &     1.4 \\
   5 & 034328.76$+$315458.9 & 55.869862 &  31.916371 &     0.1 &     7.5 &  405.0 &    21.0 &    12.0 &   241.3 &    0.89 &   18.9 & $<$-5 & .... & c &   105.7 &     2.5 \\
   6 & 034330.52$+$315649.6 & 55.877202 &  31.947123 &     0.4 &     6.6 &   32.1 &     6.6 &     4.9 &    25.6 &    0.91 &    4.5 & $<$-5 & g... & \nodata &    94.3 &     3.0 \\
   7 & 034332.08$+$320617.2 & 55.883700 &  32.104795 &     0.4 &     9.0 &  111.7 &    11.7 &    11.3 &     8.0 &    0.89 &    9.2 & $<$-5 & .... & a &    67.3 &     1.1 \\
   8 & 034336.33$+$320402.0 & 55.901402 &  32.067234 &     0.6 &     8.1 &   27.6 &     7.4 &    17.4 &    21.4 &    0.91 &    3.5 & $<$-5 & .... & b &   172.7 &     3.5 \\
   9 & 034336.47$+$320343.6 & 55.901976 &  32.062120 &     0.5 &     8.0 &   47.2 &     8.7 &    16.8 &    25.8 &    0.89 &    5.1 & $<$-5 & .... & a &   173.4 &     2.1 \\
  10 & 034342.05$+$315224.9 & 55.925209 &  31.873606 &     0.5 &     6.9 &   19.5 &     5.7 &     6.5 &    11.8 &    0.90 &    3.1 & $<$-5 & .... & a &   111.9 &     2.7 \\
\enddata

\tablecomments{Table~\ref{tab:src_properties_main} is available in its entirety in the electronic edition of A\&A.  
Interesting sources mentioned in the text are shown here for guidance regarding its form and content. 
\\Col.\ (1): X-ray catalog sequence number, sorted by RA.
\\Col.\ (2): IAU designation.
\\Cols.\ (3) and (4): Right ascension and declination (in decimal degrees) for epoch J2000.0.
\\Col.\ (5): Estimated standard deviation of the random component of the position error, $\sqrt{\sigma_x^2 + \sigma_y^2}$.  The single-axis position errors, $\sigma_x$ and $\sigma_y$, are estimated from the single-axis standard deviations of the PSF inside the extraction region and the number of counts extracted.
\\Col.\ (6): Off-axis angle.
\\Cols.\ (7) and (8): Net counts extracted in the total energy band (0.5--8~keV); average of the upper and lower $1\sigma$ errors on col.\ (7).
\\Col.\ (9): Background counts expected in the source extraction region (total band).
\\Col.\ (10): Net counts extracted in the hard energy band (2--8~keV).
\\Col.\ (11): Fraction of the PSF (at 1.497 keV) enclosed within the extraction region. A reduced PSF fraction (significantly below 90\%) may indicate that the source is in a crowded region. 
\\Col.\ (12): Photometric significance computed as net counts divided by the upper error on net counts. 
\\Col.\ (13): Logarithmic probability that extracted counts (total band) are solely from background.  Some sources have $P_B$ values above the 1\% threshold that defines the catalog because local background estimates can rise during the final extraction iteration after sources are removed from the catalog.
\\Col.\ (14):  Source anomalies: (g) fractional time that source was on a detector (FRACEXPO from {\em mkarf}) is $<$0.9.
\\Col.\ (15): Variability characterization based on K-S statistic (total band): (a) no evidence for variability ($0.05<P_{KS}$); (b) possibly variable ($0.005<P_{KS}<0.05$); (c) definitely variable ($P_{KS}<0.005$).  No value is reported for sources with fewer than four counts or for sources in chip gaps or on field edges.
\\Col.\ (16): Effective exposure time: approximate time the source would have to be observed at the aimpoint of the ACIS-I detector in Cycle ??? to obtain the reported number of net counts (see ...
\\Col.\ (17): Background-corrected median photon energy (total band).}

\end{deluxetable}

\end{document}